\documentclass[conference]{IEEEtran}
\usepackage{comment}

\usepackage{hyperref}

\usepackage{amsmath,amssymb,amsfonts}
\usepackage{amssymb}
\usepackage{algorithmic}
\usepackage{textcomp}
\usepackage{multirow}
\usepackage{enumitem}
\usepackage{textcomp}
\usepackage{makecell}
\usepackage{multirow}
\usepackage{tabularray}
\usepackage{array, makecell}

\usepackage{float}
\usepackage{fancyhdr}
\usepackage{colortbl}
\usepackage{arydshln}
\usepackage{graphicx}
\usepackage{textcomp}
\usepackage{soul}
\usepackage{hyperref}
\usepackage{tcolorbox}
\usepackage{wrapfig}
\usepackage{algorithm2e}
\RestyleAlgo{ruled} 
\usepackage{amssymb}
\usepackage{pifont}
\usepackage{amsmath}
\usepackage{graphicx}
\usepackage{textcomp}
\usepackage{makecell}
\usepackage{multirow}
\usepackage{tikz}
\usepackage{float}
\usepackage{graphicx}
\usepackage{booktabs}
\usepackage{cleveref}
\crefformat{section}{\S#2#1#3} 
\crefformat{subsection}{\S#2#1#3}
\crefformat{subsubsection}{\S#2#1#3}
\usepackage{microtype}
\usepackage{subfigure}
\usepackage{booktabs} 

\usepackage{pifont}
\usepackage{amsmath,amsfonts}
\usepackage{newtxtext,newtxmath}
\usepackage{textcomp}
\usepackage{comment}
\usepackage{algorithm2e}
\usepackage{fancyhdr}
\usepackage{lipsum}
\usepackage{comment}
\usepackage{outlines}
\usepackage{graphicx}
\usepackage{float}
\usepackage{tablefootnote}
\usepackage[export]{adjustbox}

\usepackage{multirow}
\usepackage{pifont}
\PassOptionsToPackage{dvipsnames}{xcolor}

\usetikzlibrary{shapes, arrows, positioning}
\usepackage{mdframed}

\usepackage{subcaption}

\usepackage{svg}

\usepackage{forest}

\usepackage{pifont}%
\usepackage{comment}
\usepackage{fontawesome5}

\usepackage{amsmath}

\usepackage{listings}
\def\BibTeX{{\rm B\kern-.05em{\sc i\kern-.025em b}\kern-.08em
    T\kern-.1667em\lower.7ex\hbox{E}\kern-.125emX}}

\title{Demystifying AI Platform Design for Distributed Inference of Next-Generation LLM models}

\author{Abhimanyu Rajeshkumar Bambhaniya\textsuperscript{1},  
Ritik Raj\textsuperscript{1},
Geonhwa Jeong \textsuperscript{2},
Souvik Kundu\textsuperscript{3}, \\ 
Sudarshan Srinivasan\textsuperscript{4}, 
Suvinay Subramanian\textsuperscript{5},
Midhilesh Elavazhagan\textsuperscript{4}, Madhu Kumar\textsuperscript{4},
Tushar Krishna\textsuperscript{1}
\\
  \textsuperscript{1}Georgia Institute of Technology, \textsuperscript{2}Meta, \textsuperscript{3}Intel Labs,
  \textsuperscript{4}Intel,
  \textsuperscript{5}Google
\\\vspace{2mm}
  Corresponding email: \texttt{abambhaniya3@gatech.edu}
  }

\begin{document}
\maketitle
\thispagestyle{plain}
\pagestyle{plain}

\usetikzlibrary{shadows, arrows.meta}

\tikzset{parent/.style={align=center,text width=2cm,fill=green!20,rounded corners=2pt},
    child/.style={align=center,text width=2.8cm,fill=green!50,rounded corners=6pt},
    grandchild/.style={fill=pink!50,text width=2.3cm}
}

\newcommand{\SubItem}[1]{
    {\setlength\itemindent{15pt} \item[-] #1}
}
\newcommand{\abc}{{\sf abc}\xspace}
\newcommand{\dse}{{\sf dse}\xspace}
\newcommand{\defns}{{\sf def}}
\newcommand{\tool}{{GenZ }}
\newcommand{\toolns}{{GenZ}}

\newcommand{\AB}[1]{{\color{magenta}\bfseries [Abhi:: #1]}}
\newcommand{\RR}[1]{{\color{purple}\bfseries [Ritik:: #1]}} 
\newcommand{\GJ}[1]{{\color{blue}\bfseries [Geonhwa:: #1]}} 
\newcommand{\SudS}[1]{{\color{orange}\bfseries [Sudarshan:: #1]}}
\newcommand{\SK}[1]{{\color{purple}\bfseries [Souvik:: #1]}} 
\newcommand{\TK}[1]{{\color{red}\bfseries [Tushar:: #1]}}
\newcommand{\TODO}[1]{\textcolor{red}{TODO:: #1}}
\newcommand{\rev}[1]{\textcolor{blue}{#1}}
\newcommand{\fixme}[1]{\textcolor{red}{#1 \textit{- FIXME}}}

\newcommand{\niparagraph}[1]{\vspace{1pt}\noindent\textbf{#1}}
\newcommand{\checkbox}{\textbf{\textcolor{green}{\ding{51}}}}
\newcommand{\xbox}{\textbf{\textcolor{red}{\ding{55}}}}

\newcommand*\hcircled[1]{\tikz[baseline=(char.base)]{\node[shape=circle,draw,inner sep=2pt] (char) {#1};}}

\newcommand*\circled[1]{\tikz[baseline=(char.base)]{\node[shape=circle,draw,fill=black,text=white,font=\sffamily\bfseries\small, inner sep=1.2pt] (char) {#1};}}

\def\sectionautorefname{Section}
\def\subsectionautorefname{Section} \def\subsubsectionautorefname{Section }

\def\figureautorefname{Fig.}

\definecolor{turqoise}{HTML}{4BCAD2}
\definecolor{orangered}{HTML}{CF1040}
\definecolor{redgreen}{HTML}{66AA00}
\definecolor{lightgray}{gray}{0.9}



\begin{abstract}

Large language models (LLMs) have shown remarkable performance across a wide range of applications, often outperforming human experts. However, deploying these gigantic models efficiently for diverse inference use cases requires carefully designed hardware platforms with ample computing, memory, and network resources. With constant innovation in LLM serving optimizations and model architecture evolving at breakneck speed, the hardware requirements to meet Service Level Objectives (SLOs) remain an open research question.
%
To answer the question, we present an analytical tool, \toolns, to efficiently navigate the relationship between diverse LLM model architectures, LLM serving optimizations, and AI platform design parameters. Our tool estimates LLM inference performance metrics for the given scenario. We have validated against real hardware platforms running various different LLM models, achieving a max geomean error of 5.82\%. 
We use \toolns\ to identify compute, memory capacity, memory bandwidth, network latency, and network bandwidth requirements across diverse LLM inference use cases. We also study diverse architectural choices in use today 
(inspired by LLM serving platforms from several vendors) to help inform computer architects designing next-generation AI hardware accelerators and platforms.
The source code is available at \href{https://github.com/abhibambhaniya/GenZ-LLM-Analyzer}{GitHub}. \tool can also be tried out on its \href{https://genz-llm-analyzer.streamlit.app/}{website} without any setup on your web browser.
\end{abstract}

\maketitle

\section{Introduction}
\label{sec:intro}

The success of LLMs has fueled a growing interest in Generative AI use cases - spanning Question-Answer bots, text summarization, code generation, image generation, video generation, and more. Commercial products like ChatGPT, Gemini, Github Copilot ~\cite{chatgpt, Gemini, copilot}, have performed astonishingly well in diverse metrics, often outperforming human experts ~\cite{LLMIQ}. LLMs have shown great scaling law properties~\cite{kaplan2020scaling, caballero2023broken}, with larger models ~\cite{GPT3, claude, chowdhery2022palm} demonstrating better performance as compared to smaller ones ~\cite{wei2022emergent}. 
Currently, the largest model has $\sim$1.8T parameters ~\cite{informationisbeautiful}, and future LLMs could potentially have even a few hundred trillion parameters. 

The design of LLM serving systems has become a hot area of research. This is due to its unique computational characteristics that set it apart from traditional Deep Learning inference and training. LLM serving involves two distinct stages: \textit{prefill} and \textit{decode}. The prefill stage consists of a single forward pass using all the input tokens. This is followed by an auto-regressive decode stage that generates one output token with each forward pass of the model. The prefill stage often portrays characteristics similar to a traditional forward pass (inference) and has significant computing requirements. In contrast, the decode stage consumes (and generates) one token at a time and leverages a large cache of Key and Value projections of the input tokens, requiring high memory bandwidth (BW) and capacity (especially when processing long context queries). The metrics for LLM serving are also unique - with the use case playing a key role in determining the latency criticality of prefill vs decode tokens.

\begin{figure}[!pt]
    \centering
    \includegraphics[width=\linewidth]{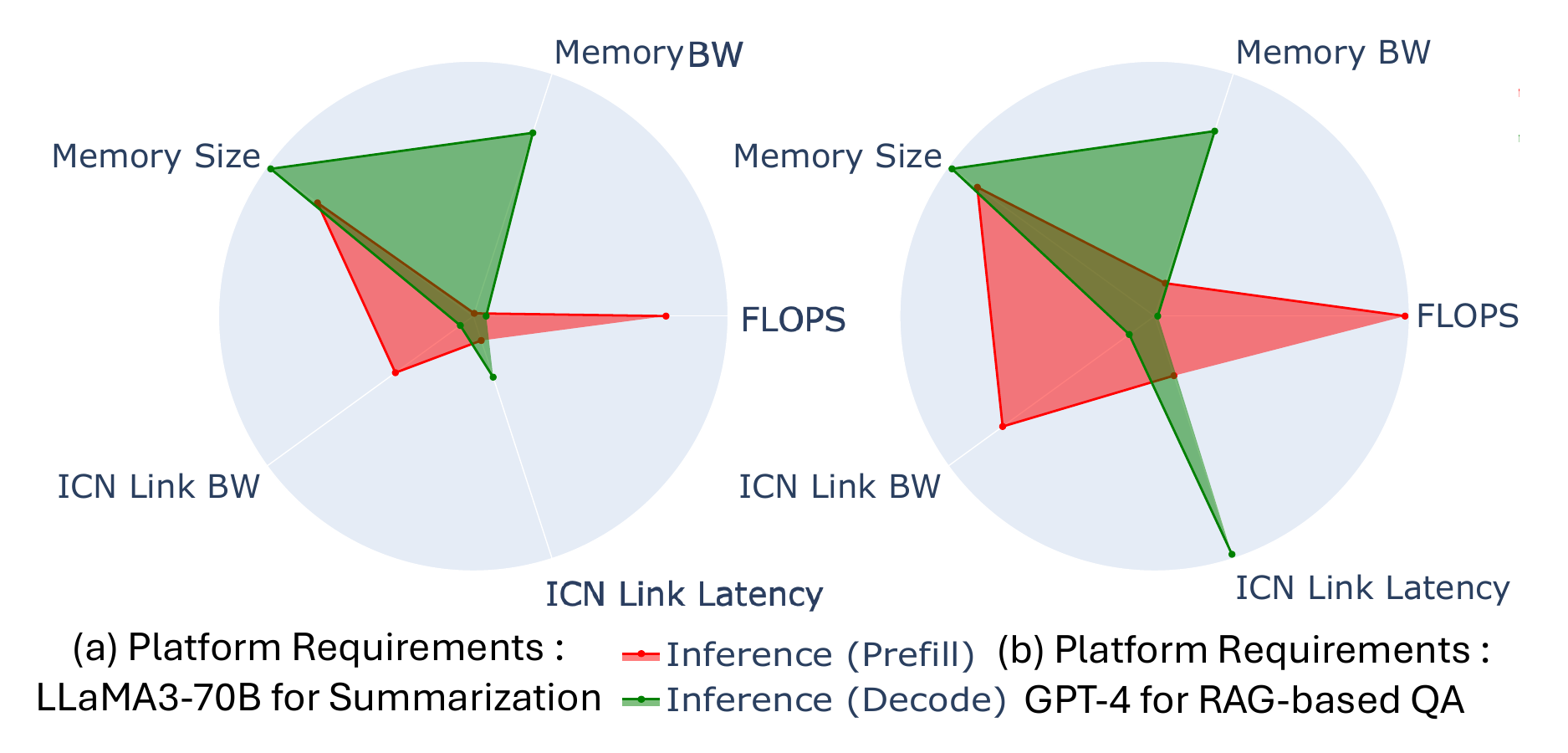}
    \caption{Platform requirements for two workloads.}
    \label{fig:various_stage_HW_req}
\end{figure}

\begin{figure*}[!t]
    \centering
    \includegraphics[width=\textwidth]
    {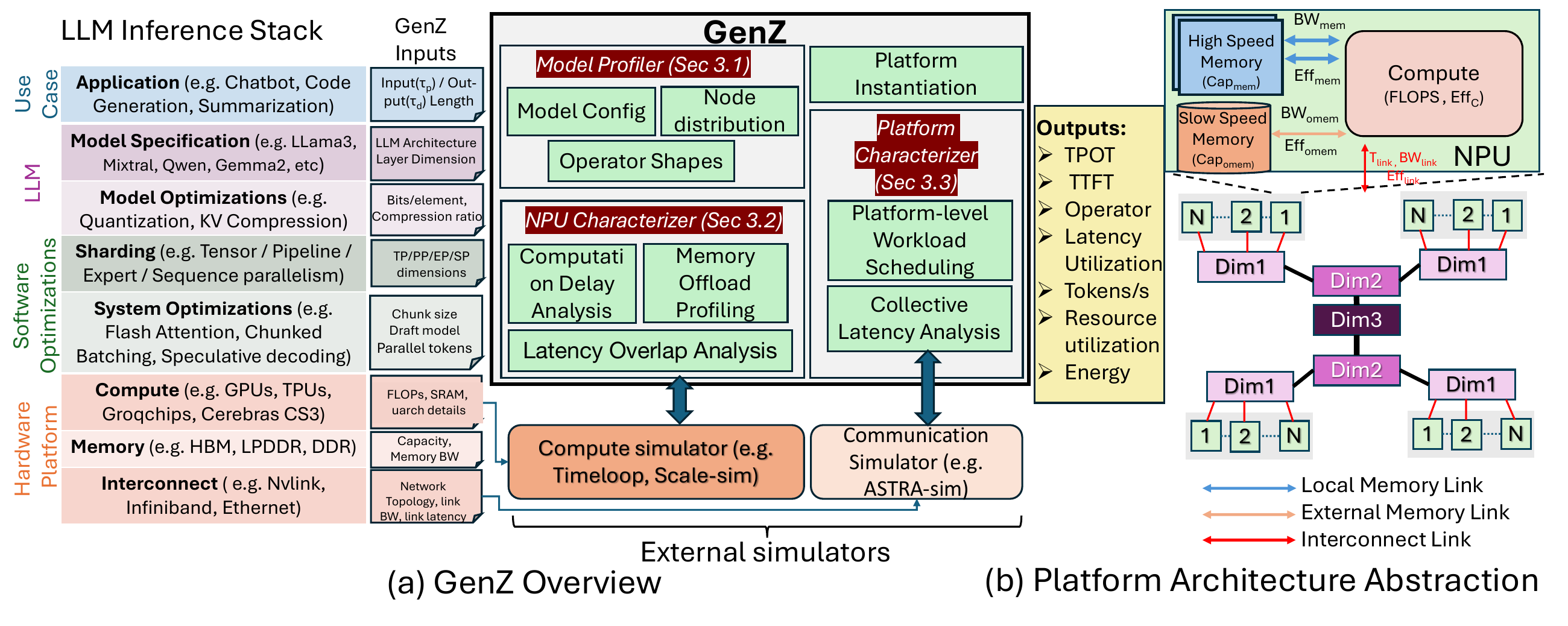}
    \caption{An overview of \tool framework. }
    \label{fig:platform_overview}
\end{figure*}



\begin{table*}[!htbp]
\centering
\resizebox{\textwidth}{!}{

\begin{tabular}{|c|c|c|ccc|}
\hline
\multirow{2}{*}{\textbf{Framework}} & \multirow{2}{*}{\textbf{\begin{tabular}[c]{@{}c@{}}LLM Architecture \\ Supported\end{tabular}}} & \multirow{2}{*}{\textbf{\begin{tabular}[c]{@{}c@{}}Model Level Optimizations* \\ System Level Optimizations$^{\dagger}$\end{tabular}}} & \multicolumn{3}{c|}{\textbf{Distributed Platform Modeling}} \\ \cline{4-6} 
&  &  & \multicolumn{1}{c|}{\textbf{Parallelism}} & \multicolumn{1}{c|}{\textbf{Network Hierarchy}} & \textbf{Memory Hierarchy} \\ \hline
ASTRA-sim ~\cite{astrasim} & \textcolor{orange}{Dense} & \textcolor{red}{None \xbox} & \multicolumn{1}{c|}{\textcolor{orange}{TP, PP}} & \multicolumn{1}{c|}{\textcolor{redgreen}{Multi-Level}} & \begin{tabular}[c]{@{}c@{}}\textcolor{redgreen}{Multi-Level (HBM } \\\textcolor{redgreen}{+ DRAM offload)}\end{tabular} \\ \hline
LLM-Viewer ~\cite{yuan2024llm} & \textcolor{orange}{Dense, Dense-GQA} & \textcolor{orange}{Flash-attention$^{\dagger}$, quantization*} & \multicolumn{1}{c|}{\textcolor{orange}{TP}} & \multicolumn{1}{c|}{\textcolor{orange}{Single Level}} & \textcolor{orange}{Single-Level (HBM)} \\ \hline
LLM-Compass ~\cite{zhang2023hardware} & \textcolor{orange}{Dense} & \textcolor{red}{None \xbox} & \multicolumn{1}{c|}{\textcolor{orange}{TP, PP}} & \multicolumn{1}{c|}{\textcolor{orange}{Single Level}} & \textcolor{orange}{Single-Level (HBM)} \\ \hline
Vidur ~\cite{agrawal2024vidurlargescalesimulationframework} & \textcolor{orange}{Dense, Dense-GQA} & \textcolor{orange}{Flash-attention$^{\dagger}$, chunked prefill$^{\dagger}$} & \multicolumn{1}{c|}{\textcolor{orange}{TP, PP}} & \multicolumn{1}{c|}{\textcolor{orange}{Real HW only}} & \textcolor{orange}{Single-Level (HBM)} \\ \hline
LLMServingSim ~\cite{cho2024llmservingsimhwswcosimulationinfrastructure} & \textcolor{orange}{Dense} & \textcolor{orange}{Flash-attention$^{\dagger}$, quantization*} & \multicolumn{1}{c|}{\textcolor{orange}{TP, PP}} & \multicolumn{1}{c|}{\textcolor{redgreen}{Multi-Level}} & \textcolor{orange}{Single-Level (HBM)} \\ \hline
\textbf{\tool (This Work)} & \begin{tabular}[c]{@{}c@{}c@{}} \textcolor{redgreen}{Dense, Dense-GQA} \\ \textcolor{redgreen}{MoE, Mamba} \\ \textcolor{redgreen}{Sliding window attention} \\ \end{tabular} & \begin{tabular}[c]{@{}c@{}c@{}}\textcolor{redgreen}{Flash-attention$^{\dagger}$, chunked prefill$^{\dagger}$, beam-search$^{\dagger}$} \\ \textcolor{redgreen}{speculative decoding$^{\dagger}$, quantization*, sparsity*} \\  \textcolor{redgreen} {mixed precision computation*, KV pruning*}\end{tabular} & \multicolumn{1}{c|}{\textcolor{redgreen}{TP, PP, EP}} & \multicolumn{1}{c|}{\textcolor{redgreen}{Multi-Level}} & \begin{tabular}[c]{@{}c@{}}\textcolor{redgreen}{Multi-Level (HBM } \\\textcolor{redgreen}{+ DRAM offload)}\end{tabular}  \\ \hline
\end{tabular}
}
\caption{Comparison of prior works for modeling LLM inference against this work. 
%
TP, PP, and EP identify as Tensor, Pipeline, and Expert parallelism, respectively.
}

\label{tab:related_works}
\end{table*}


To state that LLM inference is an active area would be an understatement. The use cases for LLM inference continue to grow by the day. This is fueled by two trends. First, new LLM models with enhanced accuracy are being released at a rapid cadence by multiple competing AI labs  ~\cite{Gemini, GPT4Arch, Claude3_7, bi2024deepseek, grok, Llama4}.
Second, for each model, a plethora of model-level (i.e., lossy) optimizations (such as quantization ~\cite{gholami2022survey, kang2024gear, xiao2023smoothquant}, pruning ~\cite{frantar2023sparsegpt,  bambhaniya2024progressive, jeong2024abstracting},
fast token decoding ~\cite{kim2024speculative,fu2024break, elbayad2019depth,chen2023ee})) and system-level (i.e., lossless) optimizations (such as paged attention ~\cite{kwon2023efficient}, flash attention ~\cite{dao2022flashattention}, chunked inference ~\cite{holmes2024deepspeed, agrawal2023sarathi}, scheduling ~\cite{yu2022orca}) are employed for enhancing performance by reducing the compute and memory footprint. Many of these optimization techniques have also become part of the popular inference-serving engine 
such as TensorRT-LLM ~\cite{TensorRT-LLM}, vLLM ~\cite{kwon2023efficient}, and DeepSpeed-FastGen ~\cite{holmes2024deepspeed}.

While GPU-based scaled-out platforms\footnote{We define platform as the complete hardware back-end including multiple NPUs (e.g., GPUs or TPUs), local memories (e.g. HBM) and inter-NPU interconnect (e.g., NVlinks/XeLinks).} have gone mainstream for large model training, the jury is out on the right architectural platform for LLM inference. Today, there exist several architecturally distinct platforms that have \textit{all} shown competitive performance across multiple LLM use cases~\cite{artificialanalysis}. Examples include GPU-based platforms from NVIDIA and AMD, programmable scaled-out dataflow processors from SambaNova, and SRAM-only architectures from Cerebras (via waferscale) and Groq (via hundreds of interconnected ASICs). There is also a lot of excitement around a plethora of emerging LLM platforms - such as transformer-specific ASICs from Etched~\cite{Etched} and high-speed photonics from Lightmatter~\cite{Photonic60:online}.

We aim to demystify and quantify design insights for AI platforms across a suite of LLM model architectures, system optimizations and use cases. %
To this end, we introduce \tool (\underline{Gen}erative LLM analy\underline{Z}er), a framework for modeling and evaluating the relationship between LLM uses cases, model optimizations, software optimizations and the hardware platform and predict the end-to-end LLM inference performance. 
It should be noted that \tool does not intend to simulate individual NPUs, rather we focus on simulating the distributed platforms for various LLM architectures combined with inference optimizations.
While there exist valuable tools in the community to design NPUs~\cite{scale-sim,timeloop,maestro} and distributed network fabric architectures~\cite{astrasim}, GenZ is the first tool, to the best of our knowledge, capturing the full-spectrum of LLM inference optimizations enabling the isolation and study of specific model/software/hardware optimizations on the end-to-end LLM performance (or energy). 
\autoref{tab:related_works} contrasts \tool against related efforts on distributed platform modeling.
In fact, GenZ is able to plug-in external tools for compute and communication to enable high-fidelity modeling of the underlying hardware.
%
%
%
We validate \tool thoroughly with various LLMs served across different platforms. Promisingly, across different workloads, \tool can closely mimic the performance on diverse real platforms, achieving a geomean error of only 5.82\%.

We demonstrate the value of \tool via multiple case studies, answering several open questions surrounding hardware requirements stemming from diverse model types, software optimizations, and architectural choices across diverse use-cases. For instance, \autoref{fig:various_stage_HW_req} highlights a subset of our key observations for two distinct 
scenarios.

In summary, this work makes the following contributions: 

\circled{1}  We introduce \toolns, an analytical framework that helps analyze LLMs combined with various inference optimizations on different platforms (\autoref{sec:methodology}). \autoref{fig:platform_overview} shows an overview of our proposed framework.
\circled{2}  Using \toolns, we study the impact of various LLM serving optimizations with current hardware specifications, and present key insights to guide the next-generation design specifications (\autoref{sec:llm_optimization_impact}).
\circled{3} We present detailed runtime analysis across various LLM architectures based on both the transformer and state-space models (\autoref{sec:model_arch_impact}).
\circled{4} Lastly, we showcase \toolns's modeling capability through four case studies on the next-generation AI inference platform design: (1) isolated scaling of different platform characteristics (\autoref{subsec:scaling_platform});
(2) comparison of different AI platform architecture design choices (for example, SRAM-Chiplet, Wafer-Scale, GPUs, ASICs) (\autoref{subsec:c2_arch_comparision}); (3) study design space choices of high bandwidth domain (HBD) size as well as associated interconnects (\autoref{subsec:c3_network_comparision}); and (4) study the impact of micro-architecture details and compute offloading(\autoref{subsec:c4_uarch_casestudy}).
\begin{figure}[!t]
    \centering
    \includegraphics[width=1\linewidth]{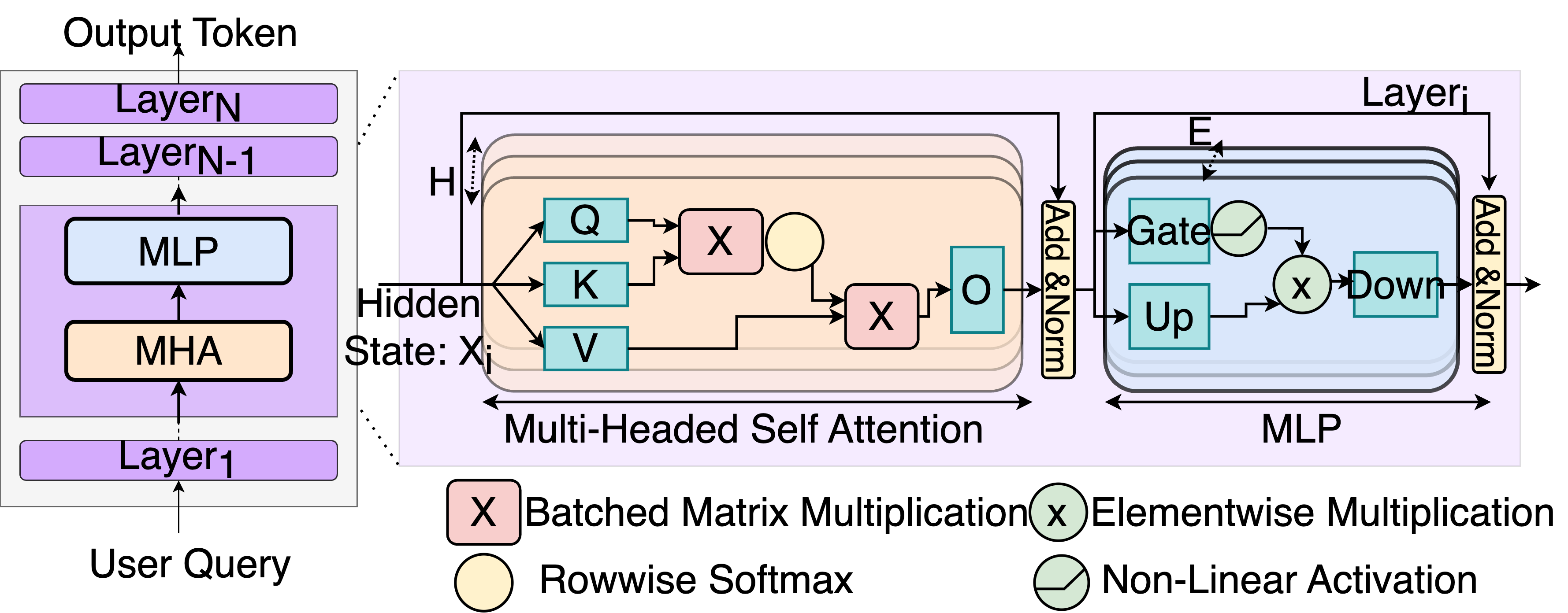}
    \caption{Typical LLM Model Architecture. Each Layer has multiple parallel heads. For MoE models, there are multiple parallel MLP layers out of which `k' are activated. 
    }
    \label{fig:transformer_overview}
\end{figure}

\section{Background}
\label{sec:backgrond}

\subsection{LLM Architecture}
\label{subsec:llm_arch}
LLMs are generally designed by stacking multiple transformer decoder layers \cite{vaswani2017attention} as shown in \autoref{fig:transformer_overview}.
Each layer includes a multi-head self-attention (MHA) and a multi-layer perception (MLP).
The key model parameters mainly include the model embedding dimensions ($D_{model}$), the number of heads (H), the feed-forward hidden dimension ($D_{ff}$), where $D_{ff}$ is  $W_{ff}*D_{model}$, and the number of decoder blocks/layers (L). 

For each decoder layer, the input sequence is projected to three linear blocks generating three activations, namely  `Query (Q)', `Key (K)', and `Value (V)'.
The Q/K/V values are then split into H chunks each of width d, where $d=D_{model}/H$ that can be computed in parallel. 
For LLMs with group query attention (GQA), $H_{kv}$ chunks of K/V are generated, and these chunks are shared by Q in $\frac{H}{H_{KV}}$ heads.
For each attention head, corresponding Q and K are fed to a batch matrix multiplication operation that is then scaled and passed through a softmax operation to compute attention scores.
The attention score is multiplied with the V chunk, generating output activations that are then projected via another linear layer.
The output of MHA is added to the input of MHA and normalized.

The MLP module consists of three linear layers. FF$_{up}$, and  FF$_{gate}$ projects the input from $D_{model}$ to the higher intermediate dimension $D_{ff}$. The output of FF$_{gate}$ is activated using non-linear operation. This activation matrix is multiplied with the output of FF$_{up}$ using an element-wise multiplication. FF$_{down}$ projects the output of element-wise multiplication back to $D_{model}$.
The output of MLP is added to the input of MLP and normalized.
The final normalized value becomes the input to the decoder layer.


Mixture of Experts (MoE)\cite{fedus2021switch} are special class of LLMs that consist of multiple ``expert" multi-layer perceptron (MLP) layers, denoted as `E', out of which `K' experts are selected for each input token. In contrast, dense language models can be considered as a special case of MoEs, where E = K = 1, meaning there is only one expert MLP layer, and it is used for every input token. By introducing multiple experts and selectively activating a subset of them for each token, MoEs can effectively scale model capacity while maintaining efficient computation and memory usage, making them a promising approach for building larger and more capable language models.
Some popular MoE models are Switch Transformers\cite{fedus2021switch}, Mixtral 8x7B\cite{jiang2024mixtral}, Mixtral 8x22B\cite{mixtral22b}, DBRX\cite{dbrx}, GROK\cite{grok}, GPT-4\cite{GPT4, GPT4Archdetails, GPT4Arch}.
\autoref{tab:model_parameters} presents some of the state-of-the-art LLMs and associated dimensions.

\begin{table}
    \centering
    \scriptsize
    \begin{tabular}{|c|c|c|c|c|c|}
    \hline
    \textbf{Model} & \textbf{$D_{model}$} & \textbf{Layers} & \textbf{\begin{tabular}[c]{@{}c@{}}Heads\\ ($H_{kv}$)\end{tabular}} & \textbf{$W_{ff}$} & \textbf{\begin{tabular}[c]{@{}c@{}}Experts\\ (Selected)\end{tabular}} \\ \hline
    Gemma2-\textbf{2B}~\cite{gemmateam2024gemma2improvingopen} & 2304 & 26 & 8(4) & 4 & - \\ \hline
    LLaMA3-\textbf{8B}~\cite{llama3} & 4096 & 32 & 32(8) & 3.5 & - \\ \hline
    Gemma2-\textbf{27B}~\cite{gemmateam2024gemma2improvingopen} & 4608 & 46 & 32(16) & 8 & - \\ \hline

    Mixtral-\textbf{8x22B}~\cite{mixtral22b} & 6144 & 56 & 48(8) & 2.66 & 8(2) \\ \hline
    LLaMA3-\textbf{70B}~\cite{llama3} & 8192 & 80 & 64(8) & 3.5 & - \\ \hline
    GPT3-\textbf{175B}~\cite{GPT3} & 12288 & 96 & 96 & 4 & - \\ \hline
    LLaMA3-\textbf{405B}~\cite{llama3} & 16384 & 126 & 128(8) & 3.25 & - \\ \hline
    GPT4-\textbf{1.8T}~\cite{GPT4} & 10752 & 120 & 84 & 4 & 16(2) \\ \hline
    Dense-\textbf{5T} & 49152 & 128 & 192(24) & 4 & - \\ \hline

    MoE-\textbf{10T} & 13824 & 128 & 108(12) & 4 & 32(4) \\ \hline
    \end{tabular}
    \caption{Different LLM architectures evaluated in this work. }
    \label{tab:model_parameters}
\end{table}

\subsection{Generative LLM Inference}
\label{subsec:gen_llm_inference}



\noindent
\textbf{Prefill.} The prefill stage runs only once on input sequence of $\tau_{p}$ tokens to generate the K and V activations, which are often kept as the KV cache, for each LLM layer. The generated KV cache would be used for all subsequent output token generation. 
%
%
The prefill stage is \textbf{mostly compute-bounded} as all input tokens can be processed in parallel.

\noindent
\textbf{Decode.} After the prefill stage, output tokens are generated auto-regressively, i.e. the last generated token is fed as an input to the LLM in each iteration, and one new output `token' is generated.
%
%
All the model weights and past KV cache are used to generate a single output token. Since input to the model is a single token, all matrix-matrix multiplications are converted into matrix-vector multiplications. This makes the generation stage \textbf{highly memory-bounded}.
At the end of each generation step, the newly generated KV cache is padded to the existing KV cache, meaning its linear growth with the output sequence length.
Generation stops when a special \textit{$<$end-token$>$} is generated or the user limits the maximum number of output tokens.
We define $\tau_{d}$ tokens are generated during the decode phase.

\textbf{Chunked Prefill} ~\cite{agrawal2023sarathi,nvidiaStreamliningInference, nvidiaDemystifyingInference, holmes2024deepspeed} or \textit{chunking} is a recent serving optimization used to reduce the hardware under-utilization by combining the two stages of LLM generation to provide better throughput.
In chunked prefill, a chunk of fixed size is fed to the model in each iteration. All existing decode batches are processed parallelly in the chunk. The remaining slots are filled by outstanding prefill requests.
Since the prefill is broken down into smaller chunks combined with the decode stage, this ameliorates the memory boundedness of the decode stage and improves the overall throughput of the system.
This often comes at slightly increased latency for the prefill stage.


\textbf{Beam search} ~\cite{sun2023allies, Kim_2022} is a heuristic search algorithm used during the decoding stage. It explores multiple potential sequences simultaneously, with the number of parallel sequences, called the beam size ($S_b$), typically set to 4.
%
%
After the common prefill stage, two parallel beams are launched during the decode phase. The beam with the highest likelihood is selected as the final output. 

\subsection{Metrics for LLM Serving}
The key metrics for LLM inference serving\cite{Agarwal_Qureshi_Sardana_Li_Quevedo_Khudia_2023} are:

\begin{enumerate}
    \item \textbf{Time To First Token} ($T_{TTFT}$): Time to complete one forward pass with entire input, $\tau_{p}$, and generate one output token.
    %
    \item \textbf{Time Per Output Token} ($T_{TPOT}$): Time to generate an output token for each request.  Consecutive tokens would have a time that is nearly identical to token generation. $T_{TPOT}$ for $i^{th}$ output token is proportional to the model weights and $\tau_{p} + i$.
    %
    %
    \item \textbf{Latency} ($T_{lat}$): The overall time it takes for the model to generate the full response for a user. Overall response latency can be calculated using the previous two metrics: $T_{lat} = T_{TTFT} + T_{TPOT} \times \tau_{d}$.
    \item \textbf{Throughput} ($\mu_{thr}$):  Refers to the number of output tokens per second an inference platform can generate over a batch size of $B$.
    $\mu_{thr}$ = $B$ / $T_{TPOT}$.
\end{enumerate}

\section{\tool: Generative LLM Analyzer}
\label{sec:methodology}


%
%


\tool is an analytical framework that can be used to study different LLM model architectures combined with the latest software optimizations on distributed current and next-generation NPU platforms. 
\tool has three key components: 1) model profiler, 2) NPU characterizer, and 3) platform characterizer.
We show an overview of \tool in \autoref{fig:platform_overview} and discuss each component in the following sections.

\begin{table}[!htbp]
\centering
\begin{tabular}{|c|c|c|c|c|}
\hline
\textbf{\begin{tabular}[c]{@{}c@{}}Real Life\\ Application\end{tabular}}& \textbf{\begin{tabular}[c]{@{}c@{}}\# of Tokens\\ Input/Output\end{tabular}} & \begin{tabular}[c]{@{}c@{}} Beam \\ Width \end{tabular} & \begin{tabular}[c]{@{}c@{}}TTFT / TPOT\\ (s) / (ms)\end{tabular} \\ \hline\hline
Question Answering & 1000/200 & 4 & 0.2 / 10\\ \hline
Chat Services & 3000/1000 & 2 & 0.2 / 10 \\ \hline
QA + RAG & 10000/200 & 4 & 0.4 / 10\\ \hline
Text Summarization  & 15000/1000 & 4 & 2 / 20\\ \hline
Code Generation & 20000/50 & 4 & 0.5 / 20 \\ \hline
\end{tabular}
\caption{Representative Use Cases of LLM models and their representative input hyperparameters.}
\label{tab:usecases}
\end{table}



\subsection{Model Profiler}
\label{sec:model_prof}

 %

\autoref{tab:model_parameters} shows the parameters of various LLM models that we study in this work. 
These models serve as a representative set of current and future LLMs.
%
%
GenZ model can model different LLM architectures, including MoE-based LLMs (Mixtral-8x7B~\cite{jiang2024mixtral} and GPT4) and Mamba-based models~\cite{gu2024mambalineartimesequencemodeling}.
LLaMA2, LLaMA3, Mixtral, and GPT3 architecture are available openly.
We represent GPT4  as a 1.8T parameter mixture-of-experts model, with 120 layers~\cite{GPT4, GPT4Archdetails, GPT4Arch}. A single layer of GPT4 has 16 experts of 111B parameters each, and two experts are activated for each token.
We also hypothesize two future LLM models~\footnote {We do not train these models for accuracy, which is not the focus of this work. We use it to study its computational behavior.} to represent scaled-up Dense-GQA and MoE models. 
%

GenZ's support for different architectures enables it to model all the SOTA LLMs.
%
For each new model, \tool uses the huggingface AutoModels~\cite{huggingface_generate_inference} to determine the exact shape of each operator in the model.
We store the operator dimension of layers.
%
Using these stored variables, we calculate the number of operations, operator execution engine, operator residency information, operator size, KV cache estimation, collective sizes, and collective groups.

Saving model operators offline allows us to profile larger context lengths for any LLM model quickly. For example, LLaMA2-7B has a context length limit of 4K, but we can extend the context length to any size using this offline modeling.

We also model popular optimizations in this stage of model profiling, including Kernel Fusion (Flash Attention~\cite{dao2022flashattention, dao2023flashdecoding, kao2021optimized}, Segment KV Caching~\cite{wu2023efficient}), model quantization, chunked prefilling~\cite{agrawal2023sarathi,holmes2024deepspeed}, speculative decoding~\cite{leviathan2023fastinferencetransformersspeculative}, and beam search~\cite{sun2023allies, Kim_2022}. Since this work focuses on LLM inference, we use FP8 model quantization for all our experiments and results unless specified otherwise.





\subsection{NPU Characterizer}
\label{sec:HW_acc_charac}


Our smallest hardware unit is the accelerator (alternatively referred as the NPU), as shown in \autoref{fig:platform_overview}. We assume each NPU has a certain number of compute cores, which can perform $FLOPS$ number of operations per second.
%
We use a variable ($Eff_C$) to account for the inefficiencies caused by the software and memory synchronization issues.
For modeling real systems, \tool uses the runtime of real systems (e.g. time to execute matmul operation on A100 GPU). This execution time is used to calculate the efficiency factor. This is the same methodology as adopted by previous works like Vidur~\cite{agrawal2024vidurlargescalesimulationframework}.
For modeling future hypothetical NPUs, the microarchitecture of NPUs plays a crucial role in determining runtime of each operator. To effectively model the effects of microarchitecture, \tool leverages external high-fidelity tools, such as Timeloop~\cite{parashar2019timeloop} and SCALE-sim ~\cite{scale-sim}, which specialize in simulating individual NPU dataflows and microarchitecture details.
\tool generates the operator dimensions for a given model architecture, system optimization, model optimization and parallelism strategy. This operator dimension is feed to the external tool to get the operator runtime, and thus get the hardware efficiency factor.
\autoref{subsec:c4_uarch_casestudy} shows a case study in which we model different microarchitecture using SCALE-sim\cite{scale-sim} and study their effect on the prefill stage of LLama-3-8B.


Each NPU provides access to two external memories (fast and slow). Faster (smaller) memory represents an HBM/DDR Memory Bank providing high BW($BW_{mem}$), while Slower (Larger) memory could be PCIe-accessible CPU or CXL-accessible SSD/Flash ($BW_{omem}$) for offload.  We also use an efficiency factor($Eff_{mem}, Eff_{omem}$) with the memory link for accurate memory access time. 

Since all operations in LLM inference have pre-determined shapes and we use all model weights and KV cache uniformly, a smart compiler can try to maximize the overlap between the operator computation and memory fetches.
Thus, we analyze the model's performance on an operator-by-operator basis. For each operator, we calculate its corresponding operations per sec ($C_{op}$) and the number of memory accesses($M_{op}$).
We follow a roofline-based approach combined with separate efficiency factors (extracted from real hardware or open source simulators) for computation FLOPs and memory BW to calculate each operator's runtime on the accelerator.

\begin{align}
    T_{op} &= max(\frac{C_{op}}{FLOPS \times Eff_{C}}, \frac{M_{op}}{BW_{mem} \times Eff_{mem}})
\end{align}

This simple yet effective modeling methodology is perfect for quickly estimating LLM serving trends on different hardware.
In \autoref{subsec:genz_validation}, we show that \textbf{using these efficiency factors, we can simulate the trends generated on real GPUs.}

\subsection{Platform Characterizer}
\label{sec:HW_platform_charac}

One of the key 
features
of \tool is its ability to simulate multi-dimensional network topologies for LLM inference.
\tool defines an inference platform as multiple NPUs connected through a multi-dimensional interconnection network (ICN), for scale-up and scale-out, as shown in \autoref{fig:platform_overview}. 
Each dimension in ICN has the following properties: link latency ($T_{link}$), network link bandwidth ($BW_{link}$), and network link efficiency ($Eff_{link}$). 

\begin{figure}[t!]
    \centering
    \includegraphics[width=\linewidth]{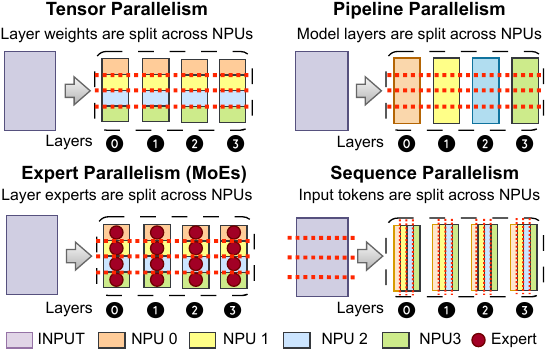}
    
    \caption{Various parallelization strategies for neural network training and inference. Each colored box represents an accelerator (NPU), and the numbers correspond to model layers. 
    }
    \label{fig:various_parallelism}
\end{figure}

\textbf{Parallelism.} Typically distributed LLMs are served using five different types of parallelism strategies:
Data Parallel (DP), Tensor parallel  (TP) ~\cite{shoeybi2019megatron}, Pipeline parallel (PP) ~\cite{huang2019gpipe}, Expert parallel (EP)~\cite{fedus2021switch} (Only for MoE models) and Sequence parallel (SP)~\cite{jacobs2023deepspeed, li2022sequence} (primarily for training long sequence).
\autoref{fig:various_parallelism} shows the example of how different parallelism splits the input tokens and model weights. 
%
%
\tool handles the overlap of physical topology and logical parallelism topology.
For our experiments, we use the parallelism order as TP:EP:PP. This order points to how the NPUs are physically located to one other. For TP:EP:PP, the NPUs doing TP are physically the closest; next, NPUs with EP, and finally, nodes with PP.
The degree of parallelism can be arbitrary, and \tool will correctly map the logical parallelism mapping on the multi-dimension physical network topology defined by the user.

\textbf{Collectives.} For each degree of parallelism, \tool generates the required collective pattern(s). There can be five kinds of collective patterns: AllReduce (TP \& EP), All-to-All (EP), Send-Recv (PP), AllGather (SP \& TP), and ReduceScatter (TP). \tool allows the all-reduce collective to be broken down into ReduceScatter followed by AllGather for hiding the communication latencies.
To get the runtime for each collective, we simulate it by calling the ASTRA-sim~\cite{won2023astra} system-layer as that provides implementations for diverse topology-aware collective algorithms.
%
%
%
In our modeling, we also have a knob to decide whether to overlap collectives with compute tasks or execute them sequentially. 
For this work, we use non-overlapping communication similar to SOTA frameworks~\cite{pagedAttention}.

\begin{figure}
    \centering
    \includegraphics[width=\columnwidth]{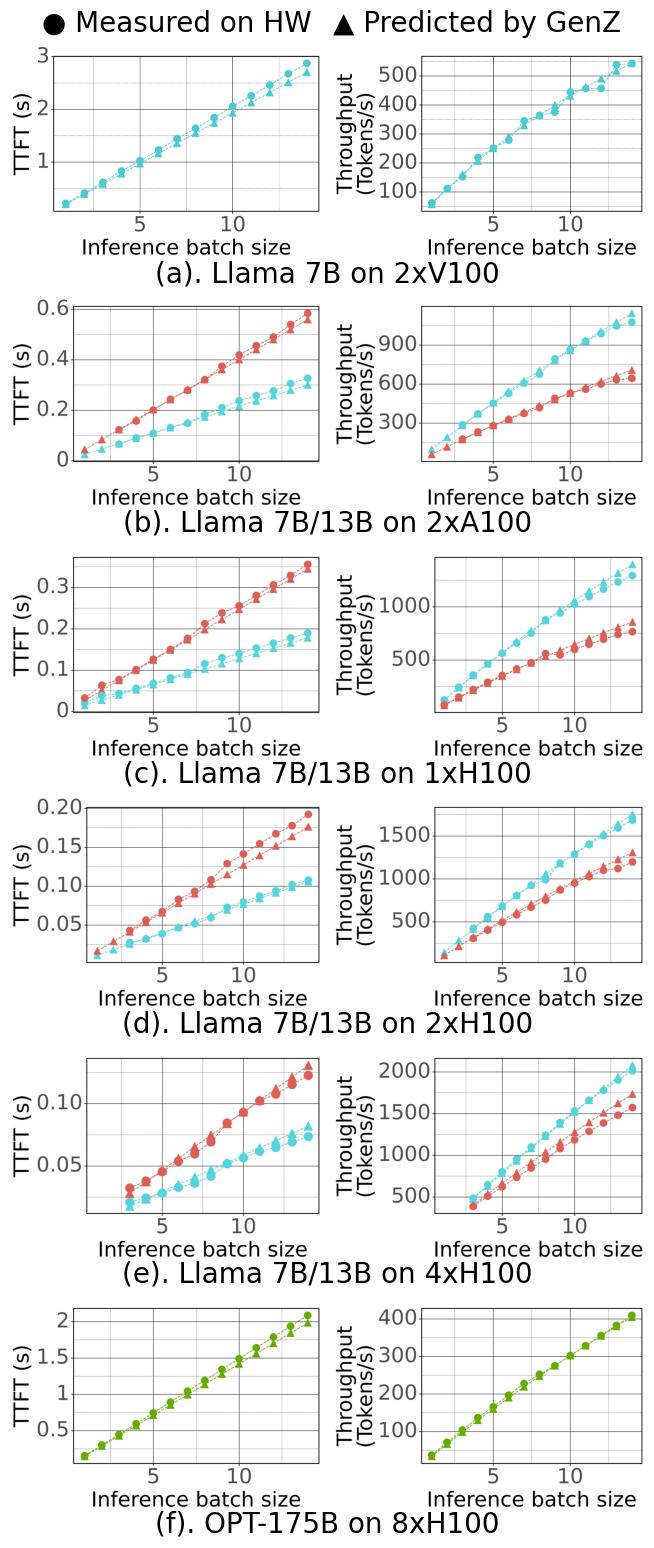}
    \caption{\tool LLM inference modeling against vLLM inference in prefill and decode stage for varying batch size on various platforms with different $\tau_{p}$. Line color  \textcolor{turqoise}{blue},  \textcolor{orangered}{red}, and  \textcolor{redgreen}{green} represent  \textcolor{turqoise}{LLaMA2-7B},  \textcolor{orangered}{LLaMA2-13B}, and  \textcolor{redgreen}{OPT-175B}. }
    \label{fig:vllm_validation}
\end{figure}


\subsection{\tool Implementation and Validation}
\subsubsection{\tool Implementation and Runtime}
\tool is implemented in over 5,000 lines of Python code and packaged as a single Python module, enabling seamless integration into design space exploration (DSE) workflows via a simple pip installation. The tool is computationally lightweight, requiring approximately 30 milliseconds per forward pass on a standard 8-core CPU, given a batch size of 64 and 1024/256 input/output tokens. Runtime efficiency is further enhanced through operator reuse: \tool identifies and skips redundant computations by sharing runtime estimates across layers.

\label{subsec:genz_validation}

\subsubsection{\tool Validation}
We validate \tool on five different real hardware platforms: a) HGX~\cite{nvidia_hgx} box with 8 NVIDIA H100 SXM GPUs (80 GB) fully connected by NVLinks, b) 2xA100, c) Intel Gaudi2, d) AMD MI300x, e) 8xSambanova SN40L.
To assess \tool’s accelerator modeling, we analyze prefill, decode, and chunking trends across these systems.

\textbf{Efficiency Factors:} Our measured efficiency factors are derived from profiling real NPUs, following a methodology similar to Vidur~\cite{agrawal2024vidurlargescalesimulationframework}. We execute the same kernel multiple times and measure average utilization to obtain realistic efficiency estimates.
LLM inference frameworks like vLLM execute static PyTorch computational graphs, ensuring consistent hardware utilization across runs. We validate our approach by comparing predicted runtimes against median measured runtimes, minimizing the impact of outliers. While we observe linearity in the validation data, it arises naturally from the predictable scaling of LLM inference workloads, where execution time for key bottlenecks (e.g., matrix multiplications, attention) scales proportionally with input size or batch size, rather than from manual tuning, as also observed in other works~\cite{chittyvenkata2024llminferencebenchinferencebenchmarkinglarge}.
For each validation study, we report the efficiency factor for the system used.

\textbf{Prefill \& Decode validation:} Using vLLM, we evaluate LLaMA2-7B, LLaMA2-13B, and OPT-175B across various platforms, using randomly generated dummy paragraphs as input, ranging from 500 to 2000 tokens, with each model generating a fixed 200 output tokens. 
 \autoref{fig:vllm_validation} compares the models' \textbf{prefill} TTFT and \mbox{\textbf{decode}} throughput on NVIDIA platform.
The geomean error in prefill and decode predictions between real and \toolns-predicted values is 2.73\% and 1.85\%, respectively, across different models and platforms. 
The average empirically-measured efficiency factors used for different hardware configurations are V100: 0.45, A100: 0.4, 1×H100: 0.55, 2×H100: 0.64, 4×H100: 0.66, and 8×H100: 0.75. 

\begin{figure}[t]
    \centering
    \includegraphics[width=\linewidth]{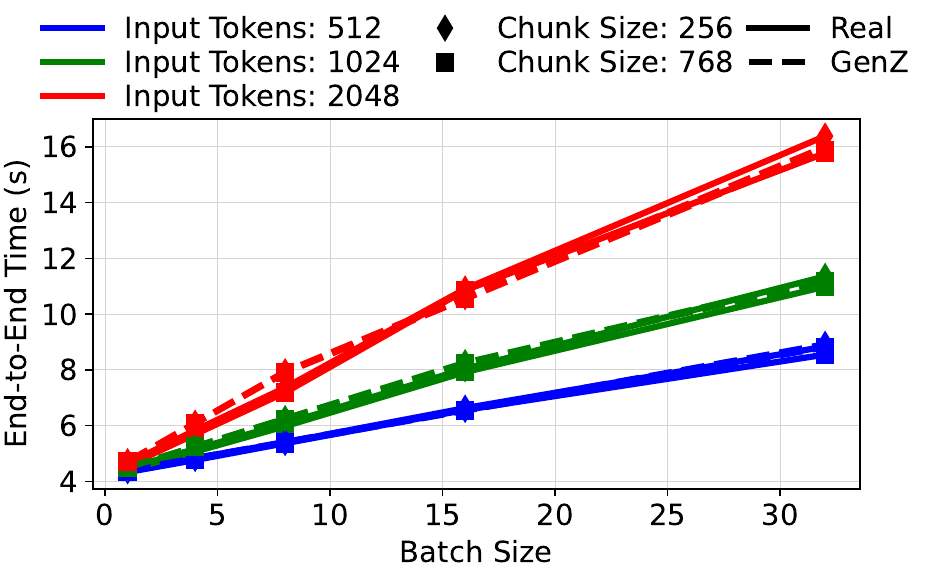}
    \caption{Validation results comparing GenZ against vLLM inference running chunked prefill on 2xA100 (Eff=0.35). Llama2-7B with varying batch sizes, input lengths, and chunk sizes.}
    \label{fig:chunking_validation}
\end{figure}

\textbf{Chunked Validation:} We run chunked serving for Llama2-7B (bf16) on 2xA100 using vLLM engine varying batch sizes (1-32), input lengths (512-2048), and chunk sizes (256,768). \autoref{fig:chunking_validation} compares actual end-to-end serving times against \tool estimates, yielding a geomean error of 1.43\%.

\begin{figure}[t]
    \centering
    \includegraphics[width=\linewidth]{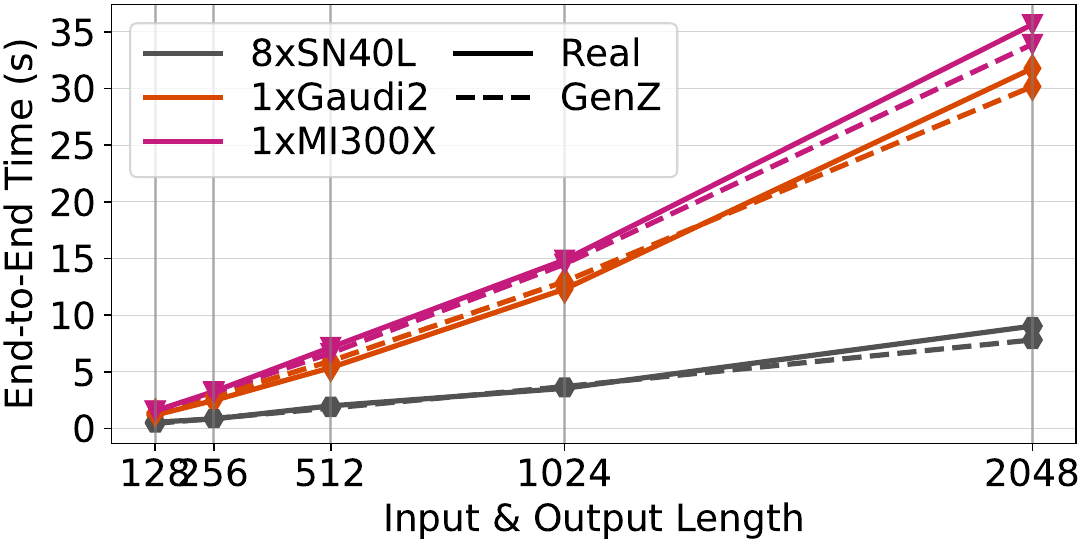}
    \caption{Validating different architectures running Llama3-8B (bf16). SN40L uses Sambaflow framework (Eff=0.9), MI300X uses vLLM (Eff=0.25) and Gaudi2 uses deepspeed (Eff=0.6). }
    \label{fig:various_arch_validation}
\end{figure}

\textbf{Validation across architectures:} We also validate \tool against three  other popular architectures\footnote{We were unable to get access to the physical node for these architecture, so we used the number from LLM-Inference-Bench~\cite{chittyvenkata2024llminferencebenchinferencebenchmarkinglarge}. The raw data was accessed from \hyperlink{https://github.com/argonne-lcf/LLM-Inference-Bench/blob/main/Plots/All_results.csv}{https://github.com/argonne-lcf/LLM-Inference-Bench}}:
(i) 8xSambanova SN40L~\cite{Sambaflow_compiler}, (ii) 1xAMD MI300X running vLLM~\cite{kwon2023efficient} and  (iii)  1xIntel Gaudiv2 running deepspeed~\cite{holmes2024deepspeed}. \autoref{fig:various_arch_validation} compares the request serving time on these platforms when running LLaMA3-8B (bf16) with batch size 16, varying input/output lengths, \tool achieves a geomean error of 5.82\% across all different architectures.

\begin{figure}[!tbph]
    \centering
    \begin{subfigure}
        \centering
        \includegraphics[width=11.5em]{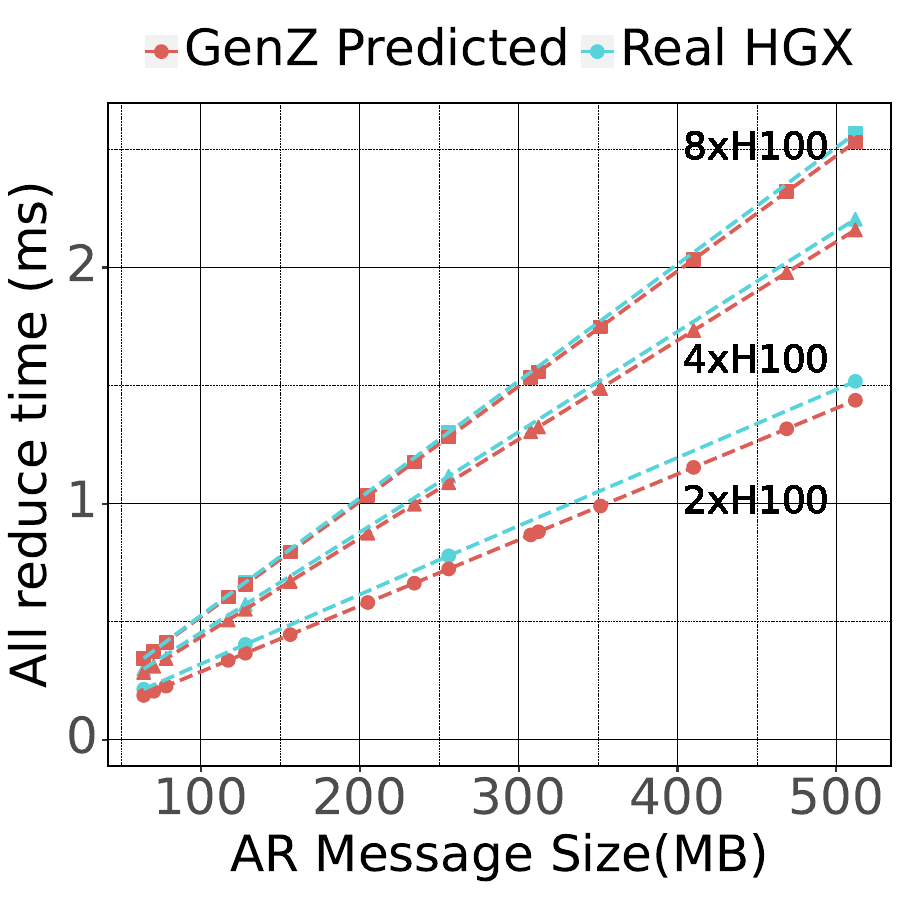}
        \label{fig:ar_prefill_validation}
    \end{subfigure}
    \begin{subfigure}
        \centering
        \includegraphics[width=11.5em]{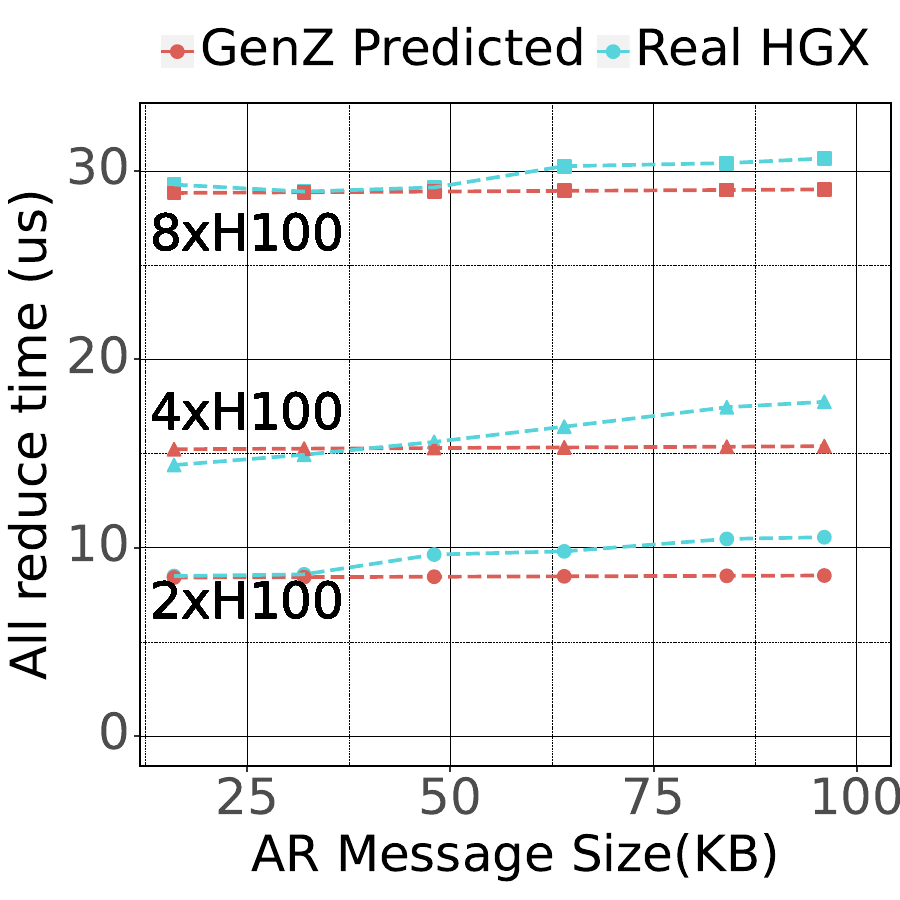}
        \label{fig:ar_decode_validation}
    \end{subfigure}
    \caption{Comparing All Reduce NCCL latency from \tool against HGX:8xH100 box. Platform sizes of 2, 4, and 8 GPUs for varying message sizes.}
    \label{fig:AR_validation.}
\end{figure}

%
\textbf{Platform interconnect validation:} To verify the collectives time at the platform scale, we benchmark all-reduce communication primitive with nccl-tests~\cite{nccl}, a communication primitive performance benchmark for NVIDIA GPUs. 
\tool's collective communication times are validated with a platform size of 2 GPUs, 4 GPUs, and 8 GPUs.
For collective communications, we observed an average link efficiency of 75\% for NVLINK, which gives an effective link BW of 350 GB/s for each GPU in HGX box.
We profile all model (\autoref{tab:model_parameters} for different input and output lengths and collect their all-reduce (AR) message sizes. 
The message size of each AR call is very small ($<$ 128 KB) for decode, while for prefill, per call message size is a few hundred MBs.
\autoref{fig:AR_validation.} compares the real hardware latency for three different platform sizes and the corresponding latency generated by \tool.
Since the prefill and decode stages have stark differences in message size, we verify the collective time for prefill and decode independently.
We found that for decode message sizes, the link latency, $T_{Link}$ is the dominating, thus the latency seems almost constant.
While prefill AR messages are large enough, that the link bandwidth, $BW_{Link}$, is the main contributor to collective time.
Collective for all datapoints, there is a 3.89\% and 2.7\% geomean error for decode and prefill message sizes between real values and \tool values.



These results confirm that \textbf{\tool accurately captures inference trends observed on real hardware.}

\section{Impact of Inference Optimizations on Hardware}
\label{sec:llm_optimization_impact}


LLM inference is one of the most active fields of research in recent years. This has led to a rapid rise in the number of innovations and optimizations being introduced. \autoref{tab:optimization_techniques} summarizes a few of the most popular techniques and their impact on compute and memory requirements from the AI inference platform. These fall into broadly 3 buckets, i.e., \circled{1} Model architecture change, \circled{2} System and algorithmic optimizations which don't change the model quality, and finally \circled{3} Algorithmic optimization with model quality changes.

\tool supports almost all of the techniques shown in \autoref{tab:optimization_techniques}, with being the \underline{only framework} that supports modeling MoEs, mamba, and hybrid models, at the same time supporting sliding window attention, speculative decoding, weight sparsity, KV pruning, and mixed precision computation and storage.

\begin{table}[!t]
    \centering
    \resizebox{\linewidth}{!}{%
    \begin{tabular}{|ccc|}
    \hline
    \multicolumn{1}{|c|}{Technique} & \multicolumn{1}{c|}{Comment} & \begin{tabular}[c]{@{}c@{}}Compute / Memory\\ Impact\end{tabular} \\ \hline
    \multicolumn{3}{|c|}{\circled{1} Foundational model architecture change (Requires pre-training)} \\ \hline
    \multicolumn{1}{|c|}{MQA/GQA} & \multicolumn{1}{c|}{Fewer KV heads} & - / \textcolor{redgreen}{↓} \\
    \multicolumn{1}{|c|}{MoE} & \multicolumn{1}{c|}{Sparesely activated FFNs} & \textcolor{redgreen}{↓} / \textcolor{orangered}{↑} \\
    \multicolumn{1}{|c|}{Sliding Window} & \multicolumn{1}{c|}{Smaller attention window} & \textcolor{redgreen}{↓} / \textcolor{redgreen}{↓} \\
    \multicolumn{1}{|c|}{Layer-wise KV sharing} & \multicolumn{1}{c|}{Multiple layers share KV cache} & - / \textcolor{redgreen}{↓} \\ \hline
    \multicolumn{3}{|c|}{\circled{2} Lossless System optimization without any impact on model quality} \\ \hline
    \multicolumn{1}{|c|}{Flash Attention} & \multicolumn{1}{c|}{Reduced memory accesses} & - / \textcolor{redgreen}{↓} \\
    \multicolumn{1}{|c|}{Chunking prefill} & \multicolumn{1}{c|}{Prefill Split + Decode} & \textcolor{orangered}{↑}  /  \textcolor{redgreen}{↓} \\
    \multicolumn{1}{|c|}{Parallelism} & \multicolumn{1}{c|}{Distributed inference} & - / \textcolor{redgreen}{↓} \\
    \multicolumn{1}{|c|}{Speculative Decoding} & \multicolumn{1}{c|}{Draft model predicts tokens} & \textcolor{redgreen}{↓}/ \textcolor{redgreen}{↓} \\ \hline
    \multicolumn{3}{|c|}{\circled{3} Lossy Model optimization with impact on model quality} \\ \hline
    \multicolumn{1}{|c|}{Quantization} & \multicolumn{1}{c|}{Reduced bit widths} & \textcolor{redgreen}{↓} / \textcolor{redgreen}{↓} \\
    \multicolumn{1}{|c|}{Weight Sparsity} & \multicolumn{1}{c|}{Removing weights} & - / \textcolor{redgreen}{↓} \\
    \multicolumn{1}{|c|}{KV pruning} & \multicolumn{1}{c|}{Removing KV cache tokens} & \textcolor{redgreen}{↓} / \textcolor{redgreen}{↓} \\
    \multicolumn{1}{|c|}{Mixed precision} & \multicolumn{1}{c|}{\begin{tabular}[c]{@{}c@{}}Different bit width for\\  storage and computation\end{tabular}} & \textcolor{redgreen}{↓} / \textcolor{redgreen}{↓} \\
    \hline
    \end{tabular}%
    }
    \caption{Various techniques for optimizing LLMs and their impact on compute and memory requirements. }
    \label{tab:optimization_techniques}
\end{table}

In this section, we use \tool to model the impact of three optimization techniques and provide insights to build next-generation AI platforms for running with those techniques.

\subsection{Chunked Prefill}

\textit{Chunked prefill}~\cite{agrawal2023sarathi} or \textit{chunking}~\cite{nvidiaStreamliningInference, nvidiaDemystifyingInference} or \textit{SplitFuse}~\cite{holmes2024deepspeed} combines the compute-bound prefill and memory-bound decode stages of LLM generation to provide better throughput. All outstanding decode batches are processed parallelly in the chunk. Additional tokens are padded by outstanding prefill requests' tokens to construct chunks of fixed size. This ensures that each forward pass always has a fixed number of tokens. Since the number of tokens in the forward pass is fixed, the runtime of most layers is always fixed. 

\begin{figure}[t]
    \centering
    \includegraphics[width=\linewidth]{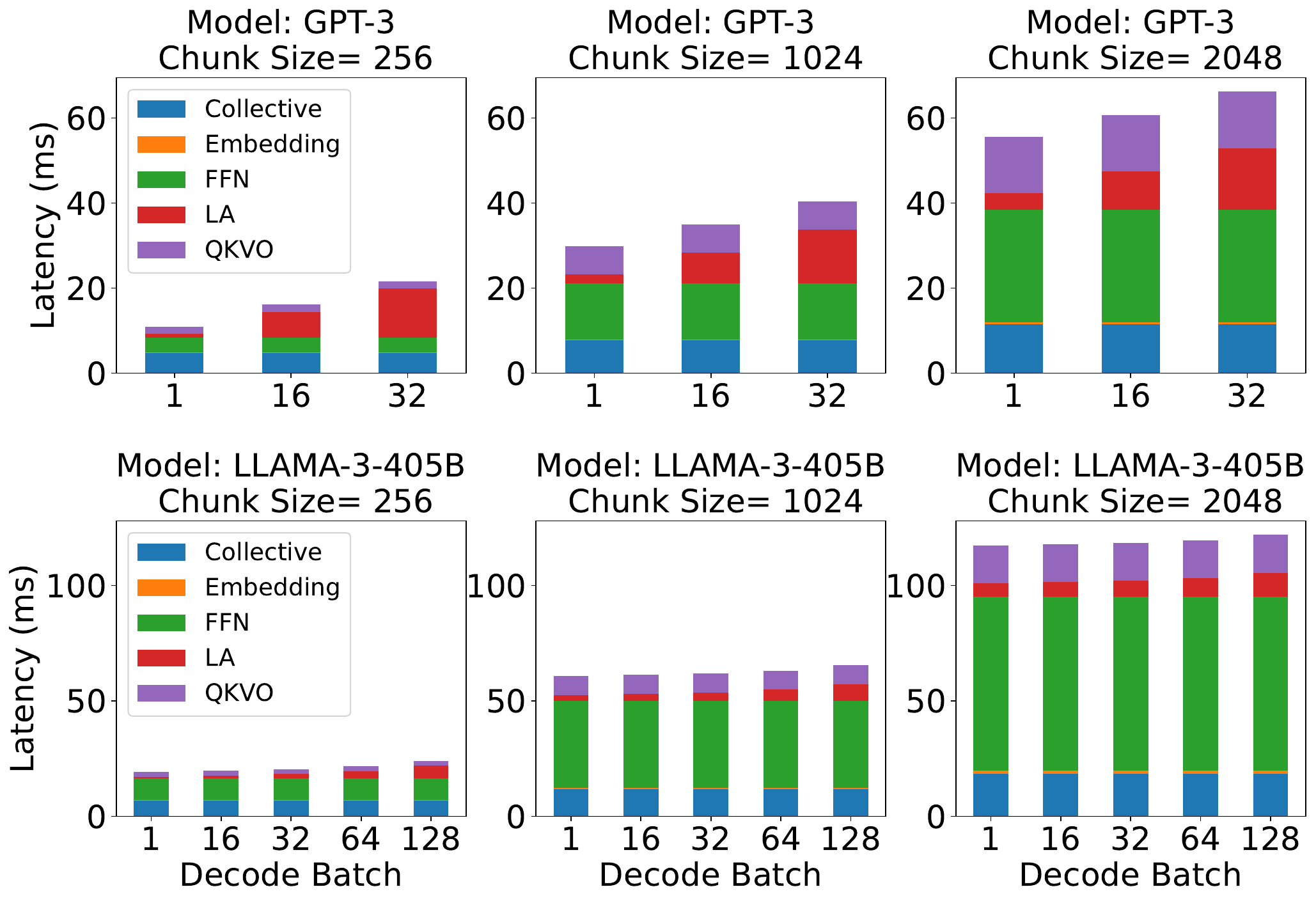}
    \caption{Runtime breakdown of inference latency with varying chunk size and decode batches.}
    \label{fig:chunking_breakdown}
\end{figure}

To understand the effect of chunking, we run two models, i.e., GPT-3 and LLama-3.1-405B, with tensor parallelism of 4 on GB200-like NPU. With $\tau_p = 4096, \tau_d=1024$. We vary the decode batch size from 1 to 128 and the chunk size from 256 to 2048.
\autoref{fig:chunking_breakdown} shows the runtime breakdown of this study. GPT-3 is unable to fit larger batch sizes as the model has dense architecture. LLama-405B, even with a larger model size, fits a much higher batch size as it uses GQA architecture.
For a given chunk size, we observe that the linear GEMM layers have nearly constant latency. Only the latency of logit and attend layers (Q.K' + Softmax + S.V) increases. This is due to the fact that these layers are also memory-bound irrespective of the context length and the batch size.
In the GPT-3 dense model, as decoded batches accumulate, the growing KV cache (red bar components) becomes a bottleneck, thus memory bandwidth becomes the bottleneck for dense models.
For the LLama-405B model with GQA architecture, the multi-headed attention(logit +attend) component is a very small part of the overall latency, and thus, latency remains didn't increase significantly with decode batch size.

\tcolorbox[width=25em, colback=lightgray,colframe=black]

  
  $\bullet$ \textbf{Memory bottleneck for dense models}: Memory BW would remain a significant bottleneck due to larger KV caches. A larger memory size would also be required to process more decode batches in parallel.

  $\bullet$ \textbf{Compute bottleneck for GQA models}: Memory critical layers contribute a very small portion of the total runtime. The compute-bound layers are the primary bottleneck for running models with GQA when doing chunking.

\endtcolorbox

\subsection{Speculative Decoding}

Speculative decoding(SD) ~\cite{leviathan2023fastinferencetransformersspeculative, kim2024speculative, DBLP:conf/emnlp/BaeKSY23, hooper2024speedspeculativepipelinedexecution, Santilli_2023, xia2023speculativedecodingexploitingspeculative} is a system level optimization technique for accelerating token generation without compromising accuracy.
It uses a smaller \textit{draft model} to generate multiple speculative token sequences in an auto-regressive fashion. These tokens serve as "guesses" for the large target model.
The large target model evaluates these guesses in a single pass. It either accepts the suggested tokens if they align with its probability distribution or rejects them and adjusts the next output accordingly.
%
%
On the flip side, if any token generated by the draft model is rejected by the target model, all subsequent tokens are dropped. Thus, the overall output throughput would be reduced.
%

We use two hyperparameters for modeling SD in \tool: i) N = number of tokens generated by the draft model before the large full model checks them. ii) $\gamma$ = probability of each token generated by the smaller model being accepted by the target model.
This helps us estimate the expected tokens per pass of the target model as 
%

$E[T] = \sum_{i=1}^{N-1} i \cdot \gamma^i \cdot (1 - \gamma) + N \cdot \gamma^N$

\begin{figure}[!t]
    \centering

\begin{tikzpicture}[scale=1]  

\definecolor{encoderColor}{RGB}{255,100,50}   
\definecolor{encoderdraftColor}{RGB}{255,255,0}

\definecolor{decoder1Color}{RGB}{100,30,230}  
\definecolor{decoder2Color}{RGB}{100,149,237}

\draw[fill=encoderColor] (0,0) rectangle (1,0.4);
\foreach \x in {1, 1.3, 1.6, 1.9, 2.2, 2.5, 2.8, 3.1, 3.4, 3.7, 4, 4.3, 4.6, 4.9, 5.2} {
    \draw[fill=decoder1Color] (\x,0) rectangle (\x+0.3,0.4);
}

\draw[fill=encoderColor] (0,0.8) rectangle (1,1.2);
\draw[fill=encoderdraftColor] (1.2,0.8) rectangle (1,1.2);
\foreach \x in {1.2,1.75, 2.3, 2.85, 3.4} {
    \draw[fill=decoder1Color] (\x,0.8) rectangle (\x+0.55,1.2);
    \foreach \y in {0.05,0.10,0.15,0.20,0.25} {
        \draw[fill=decoder2Color] (\x+\y-0.05,0.8) rectangle (\x+\y,1.2);
    }
}

\node[left] at (-0.2,0.2) {\small Base};
\node[left] at (-0.2,1.0) {\small $N=5$};
\node[below] at (2.5,-0.2) {\small Wall time $\rightarrow$};

\begin{scope}[shift={(5.6,0)}]
    \draw[fill=encoderColor] (0,1.05) rectangle (0.5,1.35);
    \node[right,font=\small] at (0.5,1.2) {Target pre.};

    \draw[fill=encoderdraftColor] (0,0.65) rectangle (0.5,0.95);
    \node[right,font=\small] at (0.5,0.8) {Draft pre.};
    
    \draw[fill=decoder1Color] (0,0.25) rectangle (0.5,0.55);
    \node[right,font=\small] at (0.5,0.4) {Target dec.};
    
    \draw[fill=decoder2Color] (0,-0.15) rectangle (0.5,0.15);
    \node[right,font=\small] at (0.5,0.0) {Draft dec.};
\end{scope}
\end{tikzpicture}
\caption{Comparing baseline LLM inference against inference using speculative decode execution with a draft model and a number of parallel decode tokens, $N=5$.}
\label{fig:specdecode_timeline}
\end{figure}

\begin{figure}[!t]
    \centering
    \includegraphics[width=1\linewidth]{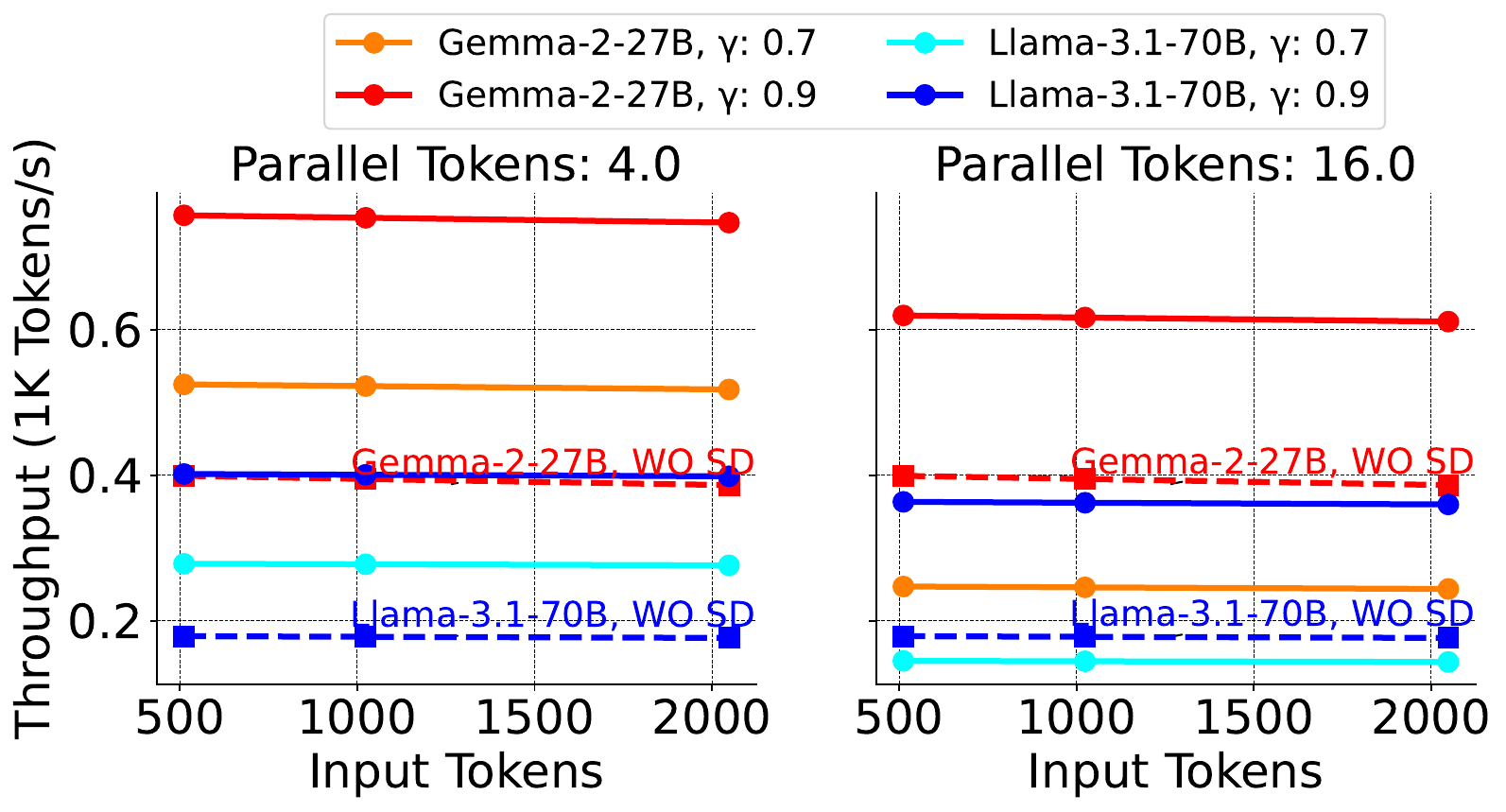}
    \caption{Decoding Throughput with speculative decoding.}
    \label{fig:spec_decode_trends}
\end{figure}

\autoref{fig:spec_decode_trends} shows decode throughput on GB200-like NPU with TP=2. We test with $N \in \{4, 16\}$ \& $\gamma \in \{0.7, 0.9\}$ for Llama-3.1-70B (draft model: Llama-3.1-8B) 
and Gemma-2-27B (draft model: Gemma-2-2B). We also vary the $\tau_p, \tau_d \in \{512, 1024, 2048\}$. The dashed lines represent the model throughput without SD. \circled{1} We see that as the number of parallel tokens, N, increases, the throughput goes down. This is because of the increased cost of running the draft model. For $\gamma = 0.7$ \& N=16, both the models perform worse than they would without SD. \circled{2} For a given number of parallel decode tokens, i.e., fixed N, $\gamma$ is directly linked to the throughput gains. We see that for N = 4, $\gamma = 0.5$ has roughly the same throughput as the model w/o SD.

The draft models, Gemma-2-2B and Llama-3.1-8B, require 10.8\% and 9.6\% extra memory for weights and 40\% and 28\% extra memory for KV cache. For a batch of 4 requests with input/output length 32K, the total additional memory requirements would be 24.7\% and 28.2\%, respectively.

\tcolorbox[width=25em, colback=lightgray,colframe=black]
  
  $\bullet$ \textbf{Larger memory capacity}: Keeping 2 models on-device with their corresponding KV cache requires HW with larger memory capacity compared to running a single model.

  $\bullet$ \textbf{Compute bottleneck}: With multiple parallel tokens being fed to the target model, most operators can be pushed to the compute-bound region. This means that more layers are compute-bound than memory-bound.
\endtcolorbox

\subsection{Optimal Parallelization Strategies for MoEs}
\label{sec:parallelism_strategies}

%
%
%
%

\begin{figure}[htpb!]
    \centering
    \includegraphics[width=\linewidth]{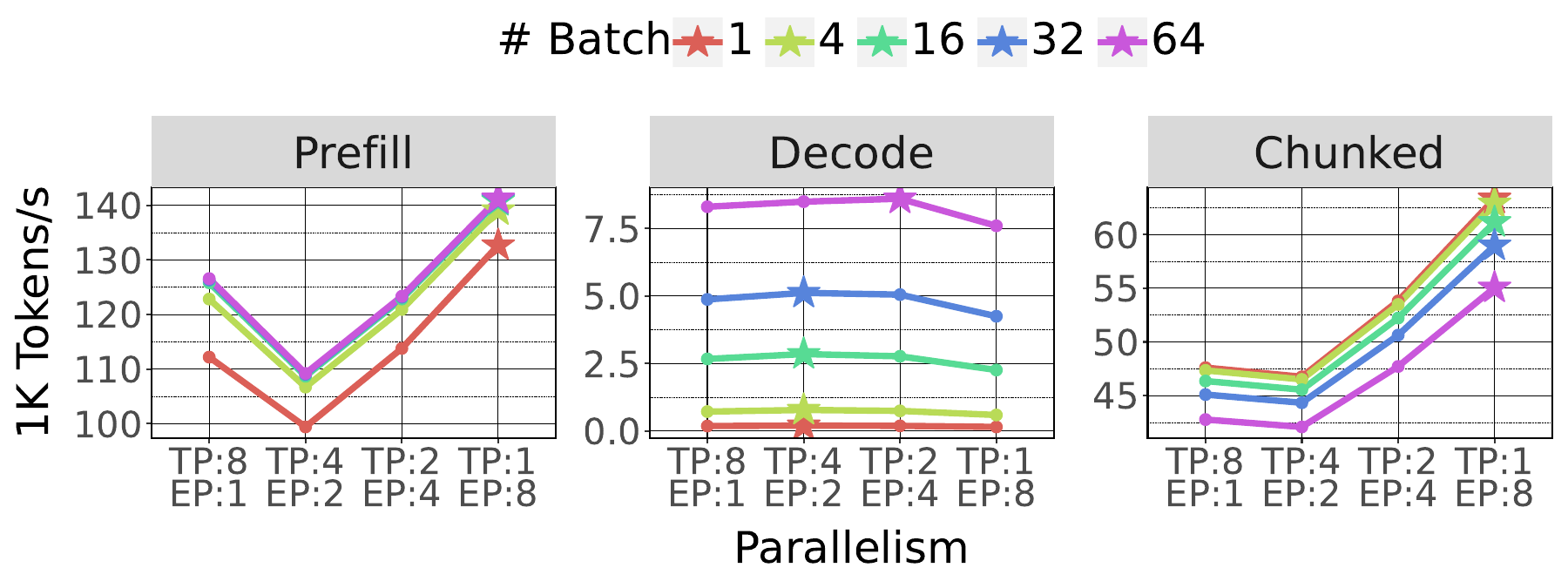}
    \caption{Comparing different parallelism strategies for Mixtral-8x22B model inference on HGX:H00x8.  $S_b = 4, \tau_p = 4k, \tau_d=256, Chunk=512$. }
    \label{fig:parallelism_strategy_comp}
\end{figure}

While tensor parallelism is generally known to be the best parallelism type for dense LLMs during inference~~\cite{chittyvenkata2024llminferencebenchinferencebenchmarkinglarge}, it is still unclear which parallelism technique would work well with MoE models.
Using \tool, we explore different parallelism strategies for a popular MoE, Mixtral-8x22B, on an H100x8 node connected by a switch network shown in \autoref{fig:parallelism_strategy_comp}. We assume that tokens are distributed among the experts in a balanced fashion for prefill. \footnote{Exploring the effect of load imbalance among experts is left as future work.} However, the small number of tokens makes it very unpredictable during the decode. Thus, TPOT of Mixtral 8x22B on 4 H100 with expert parallelism can vary between 3.23 ms (All tokens distributed equally) and 11.33 ms (All tokens going to a single expert) for batch 32.
Although different scenarios and assumptions could result in choosing different parallelism as optimal, we believe our tool, \tool can be used to find optimal parallelism for future MoEs on any HW platforms.
%
%
%
%

\tcolorbox[colback=lightgray,colframe=black]
  
  $\bullet$ \textbf{EP is preferred when all experts activated}: For prefill and chunked stages, \textit{load balanced} expert activation, EP is the best parallelism strategy. In case there is an imbalance among experts, EP could perform significantly worse.
 
  $\bullet$ \textbf{Mix of TP and EP is preferred when partially activated experts}: For the decode stage, where only a subset of experts are activated with very few tokens being routed to each expert, TP only or TP + EP is generally superior for throughput.
\endtcolorbox

%
\section{Impact of Model Architectures on HW Scaling}
\label{sec:model_arch_impact}

LLM architecture plays a pivotal role in determining their computational efficiency, scalability, and inference behavior. Model architectures are constantly evolving to optimize for specific challenges, such as memory usage, computational cost, and performance across diverse tasks. We compare the four most prominent LLM architectures, covering all SOTA open-source released models. 

\textbf{Traditional Transformers (Dense)}: 
Original foundational models like Transformer~\cite{vaswani2017attention}, GPT, and BERT have a fully connected attention mechanism where every token attends to every other token.
This architecture have quadratically scaling compute and memory  with input sequence length. 
\begin{figure}[!bt]
    \centering
    \subfigure[Context $\rightarrow$ TTFT.]{\includegraphics[width=0.24\textwidth]{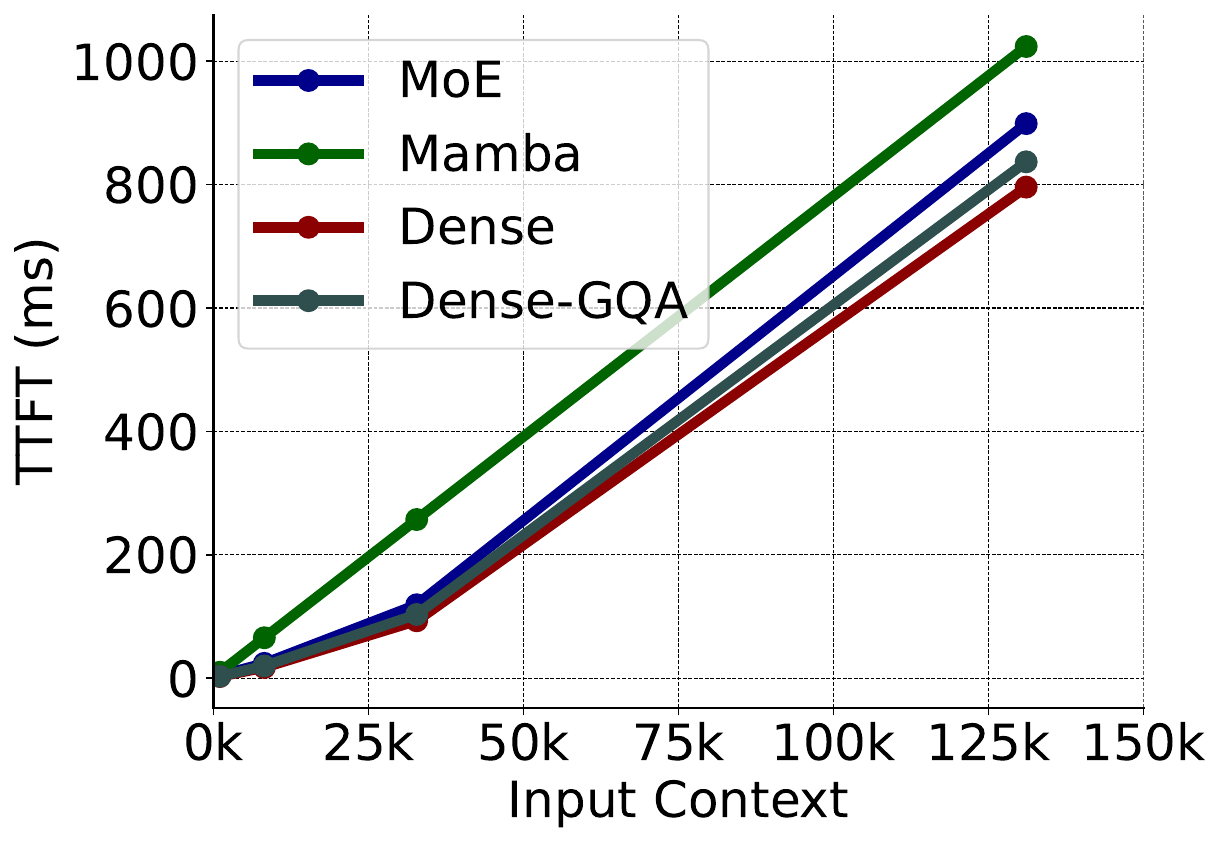}} 
    \subfigure[Batch size $\rightarrow$ TTFT.]{\includegraphics[width=0.24\textwidth]{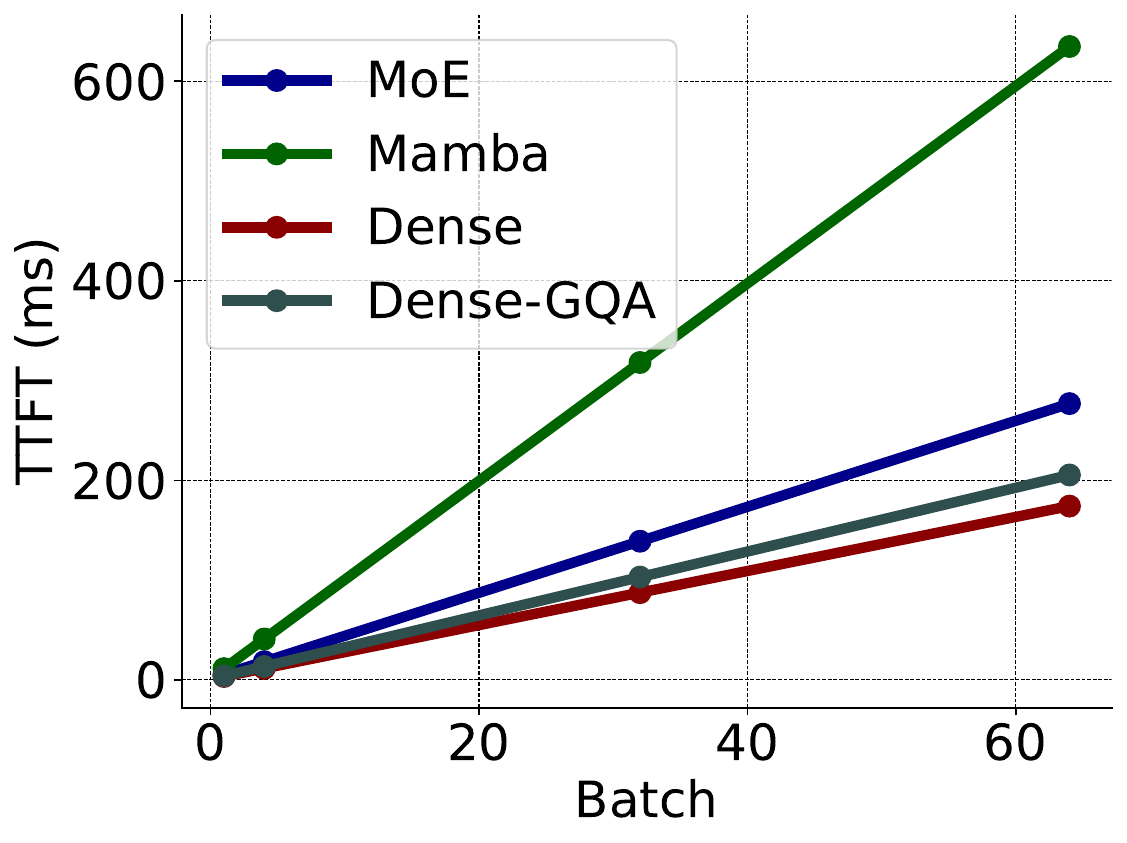}}
    \subfigure[Context $\rightarrow$ TPOT.]{\includegraphics[width=0.24\textwidth]{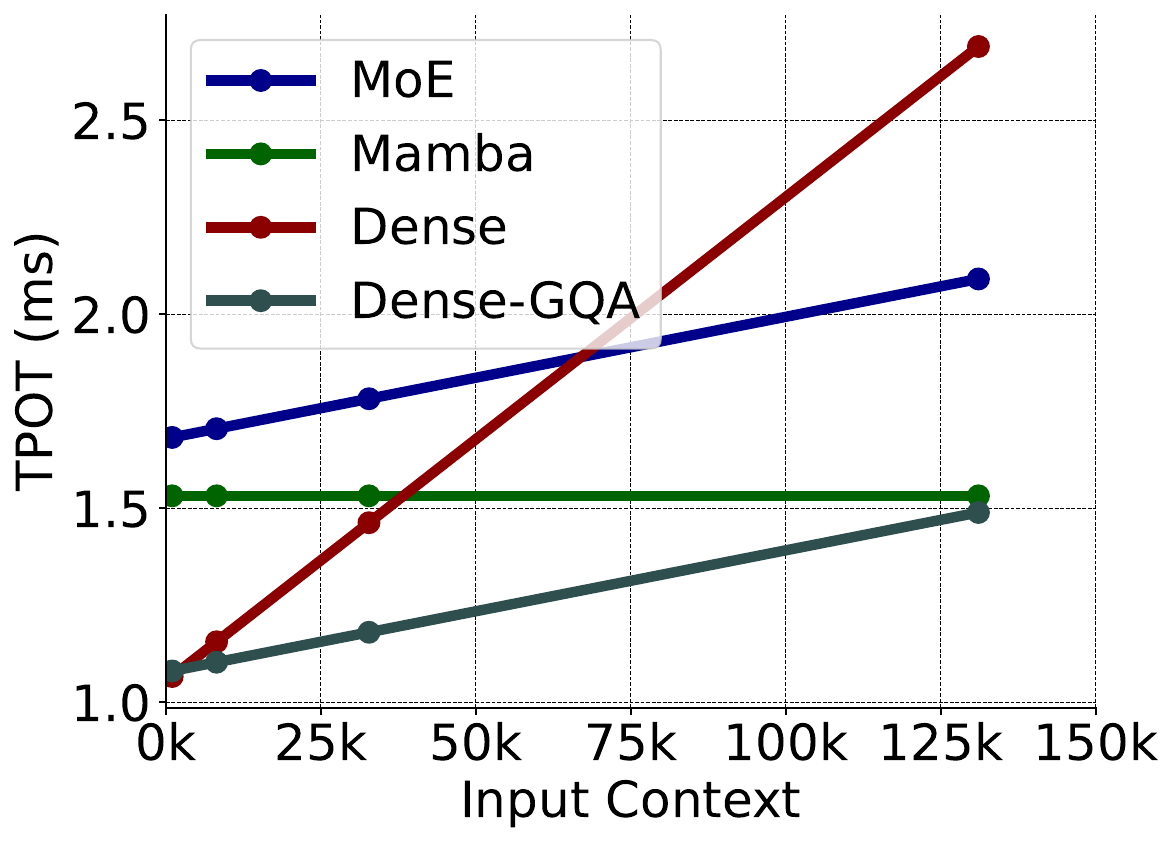}} 
    \subfigure[Batch size $\rightarrow$ TPOT.]{\includegraphics[width=0.24\textwidth]{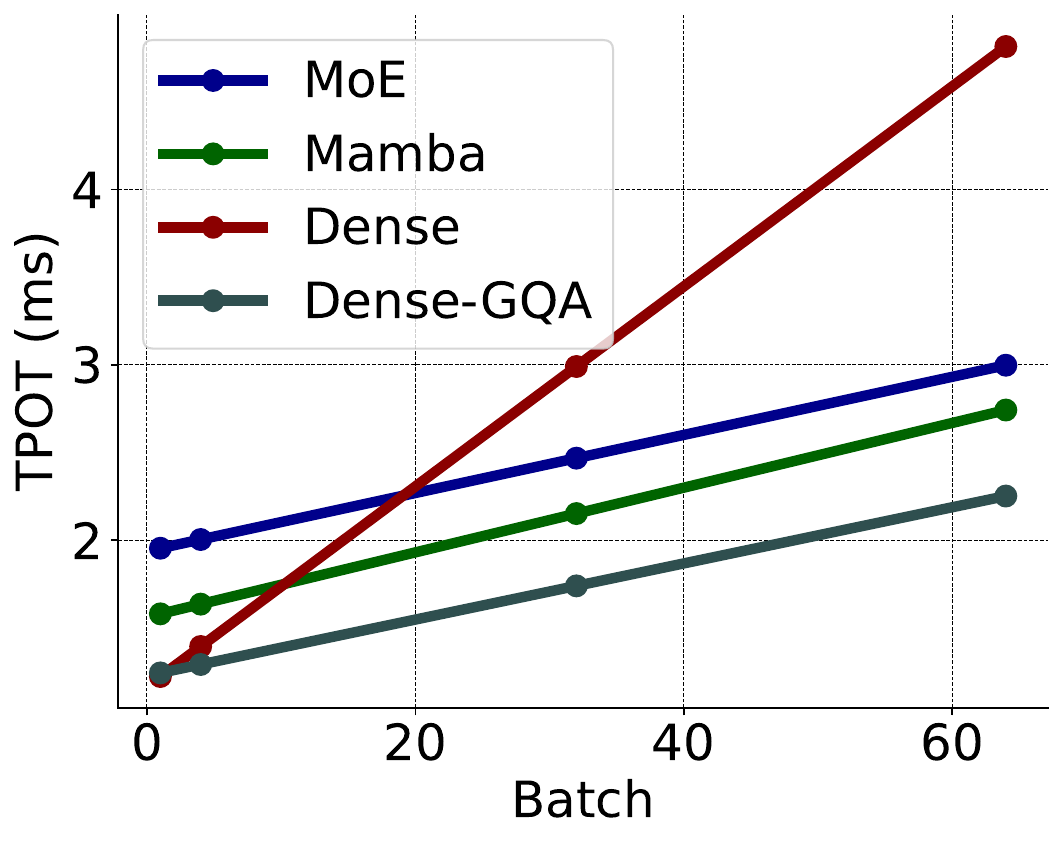}}
    \subfigure[Context $\rightarrow$ chunk time]{\includegraphics[width=0.24\textwidth]{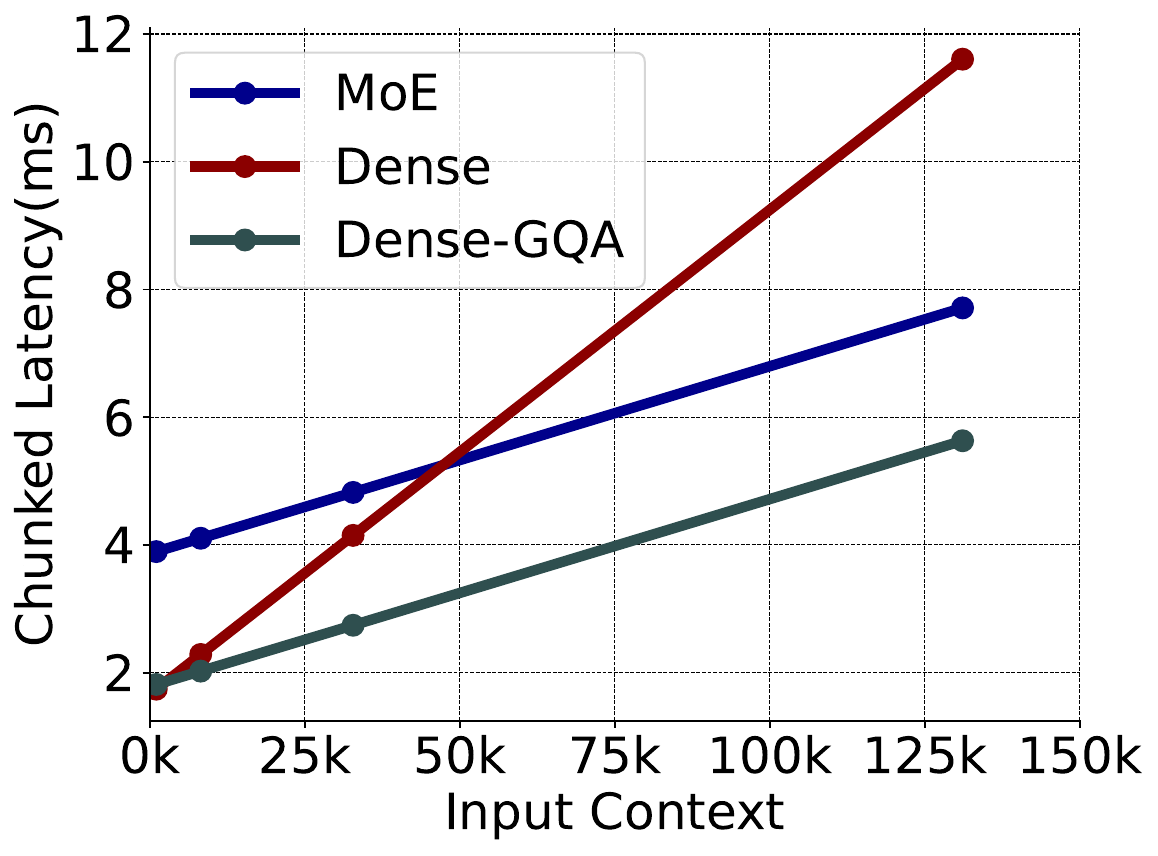}} 
    \subfigure[Batch $\rightarrow$ chunk time]{\includegraphics[width=0.24\textwidth]{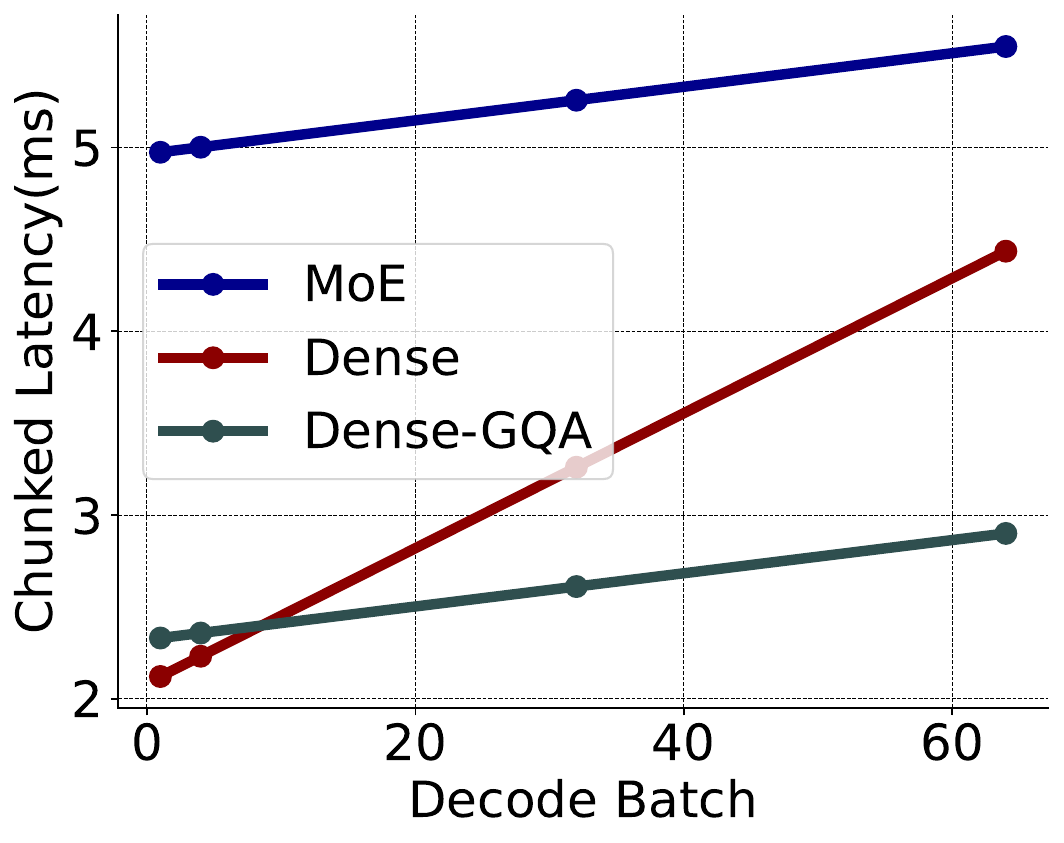}}
    \caption{Effects of context length and batch size on different model architectures during different stages.}
    \label{fig:model_arch_impact}
\end{figure}
%

\textbf{Dense Transformers with Group-query Attention}: 
Group-query attention (GQA)~\cite{hudson2019gqa} mitigates rising memory costs by sharing key-value caches across heads. This optimizes attention sparsity while maintaining accuracy, improving efficiency for long-context tasks.

\textbf{Mixture of Experts (MoE)}:
MoEs leverage sparsity by activating only a subset of "expert" sub-networks for each token.

\textbf{Mamba}:
The Mamba architecture
combines attention mechanisms with flavor of RNNs. Mamba focuses on optimizing memory and compute utilization, making it particularly advantageous for large-scale deployments.

To understand the impact of these model architectures in the various stages of LLM inference, we compare a model of each type of comparable size. We pick LLama2-7B ~\cite{touvron2023llama2}, LLama-3.1-8B~\cite{llama3}, Mixtral-8x7B (12.8 Active) ~\cite{jiang2024mixtral}, and Falcon-Mamba-7b ~\cite{zuo2024falconmambacompetitiveattentionfree} for our comparison. We study the impact of increasing context length \& batch size on the latency of stages.

\circled{1} \textbf{Impact on prefill stage}: As the context grows (\autoref{fig:model_arch_impact}(a)), All models exhibit linear scaling, for mamba model even though cache size is constant, they still need to have initial scan step linearly. With custom implementation of scan kernels, growth could be made sub-linear ~\cite{gu2024mambalineartimesequencemodeling}. For growing batch sizes (\autoref{fig:model_arch_impact}(b)), all architectures show a linear increase in latency. Dense, Dense-GQA, and MoE scales maintain more efficient scaling. We don't see any particular benefit in the Dense-GQA since the model is mainly compute-bounded, and having fewer KV heads doesn't have any effect on the runtime.

\circled{2} \textbf{Impact on decode stage}: 
For increasing context lengths (\autoref{fig:model_arch_impact}(c)), Mamba's runtime doesn't change since the generation time is context length independent. Dense sees a significant rise as context length increases due to the quadratic cost of dense attention. The effect on dense-GQA and MoE is mitigated since the KV cache growth factor is much smaller.
Growing batch size (\autoref{fig:model_arch_impact}(d)), Dense-GQA, mamba, and MoE grows almost linearly, which shows batching can help in increasing the throughput. For the dense model, the growth is linear, but the slope is higher due to the larger size of the KV cache.

\circled{3} \textbf{Impact on chunked stage}: In chunked prefill, a chunk of 512 tokens is constructed with multiple decode batches. Rest of the tokens are taken as a small chunk from the prefill request. Increasing context lengths for the decode batches,(\autoref{fig:model_arch_impact} (e)), similar to the decode stage, larger context length equates to larger KV caches, and thus latency increases linearly. 
For batch sizes scaling (\autoref{fig:model_arch_impact} (f)), Dense-GQA and MoE have a low growth rate due to fewer KV heads leading to smaller cache size. 
An interesting observation is that the MoE model has a larger latency compared to the dense model since the prefill portion of the chunk will ensure all experts are activated and thus would lead to higher latency.
\section{Platform Requirements Estimation}
\label{sec:platform_requirements}
In this section, we answer the question that \textbf{given a LLM use case, what should be the platform configuration to meet the SLO requirements?}
Prefill and decode phase presents its own set of unique platform requirements.
The key platform requirements terms are computation operations (PetaFLOPs), memory bandwidth (TB/s), and memory capacity (GBs). These metrics are crucial for understanding the hardware resources needed to run LLM models on various workloads efficiently.
These requirements are dictated by \textit{workload}.
Our aim is to quantify these requirements and provide analytical equations to scale the requirements.
We study each of the three requirements in isolation by assuming the rest of the components are not the bottleneck for the use case and model.




\begin{figure*}[t!]
    \centering
    \includegraphics[width=1\linewidth]{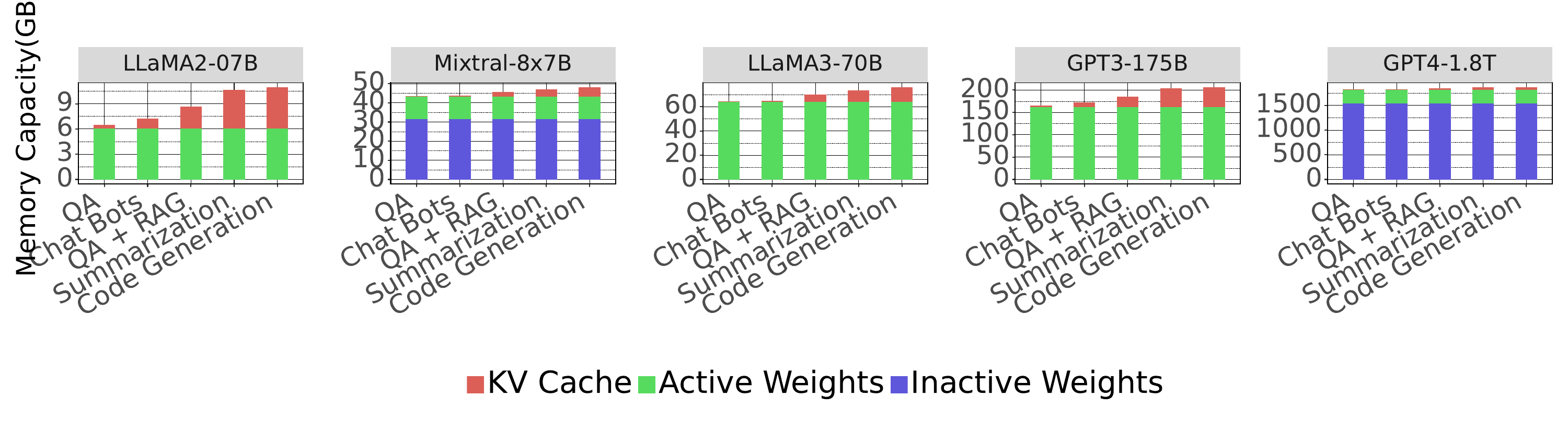}
    \caption{Memory Capacity (GB) required for various models on different use cases during the decode stage. }
    \label{fig:memory_cap_req}
\end{figure*}

\subsection{Platform Memory Capacity Requirement}
\label{subsec:platform_mem_req}

For LLM inference on the platform, there should be enough memory capacity to store complete model parameters + KV cache in the fast memory of NPUs. 
Where KV cache $= 2*B*(\tau_p + S_b*\tau_d)*H_{kv}*D/H*N_{layers}$.

Figure \ref{fig:memory_cap_req} illustrates the distribution of memory requirements across various models and usage scenarios. For LLaMA2-7B, LLaMA3-70B, and GPT3-175B, all model parameters are utilized in each inference iteration, whereas in Mixtral-8x7B and GPT4-1.8T, only 27\% and 15\% of the total model parameters are activated per inference cycle.

As model sizes increase, the ratio of KV cache parameters to active weights diminishes progressively. Specifically, the ratio of the largest KV cache size (Code Gen) to active weights is 82\%, 11\%, 20\%, 27\%, and 2.8\% for LLaMA2-7B, Mixtral-8x7B, LLaMA2-70B, GPT3-175B, and GPT4-1.8T, respectively.
Moreover, MoE models exhibit significantly smaller KV caches compared to dense LLMs. Additionally, it is noteworthy that the KV cache expands linearly with the batch size. 
Furthermore, Mixtral 8x7B and LLaMA3-70B employ GQA to minimize the KV cache footprint, thus enhancing batch size capabilities.

\vspace{1em}

\noindent\fbox{%
  \parbox{25em}{
  \textit{Key Takeaways:} \\
    $\bullet$ Fixed-usecase, MEM-CAP$_{Req} \propto O(Model Size + KV cache)$. \\
    $\bullet$ Fixed-model, MEM-CAP$_{Req} \propto O(B * (\tau_p + S_b * \tau_d))$.
  }%
}

\begin{figure}[!ht]
        \centering
    \begin{subfigure}
        \centering
        \includegraphics[width=23em]{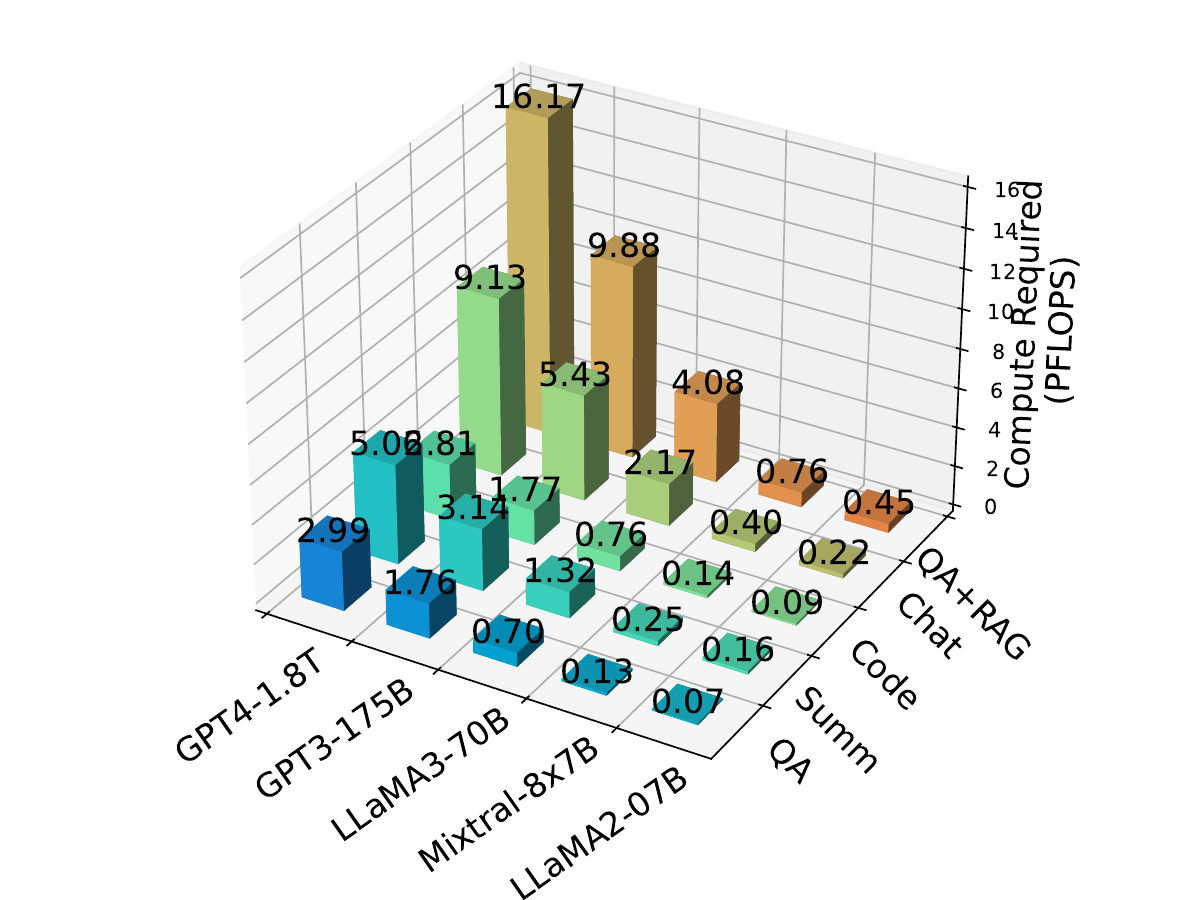}
        \text{(a) Compute PFLOPS requirement.}
        \label{fig:compute_req}
    \end{subfigure}
    \begin{subfigure}
        \centering
        \includegraphics[width=23em]{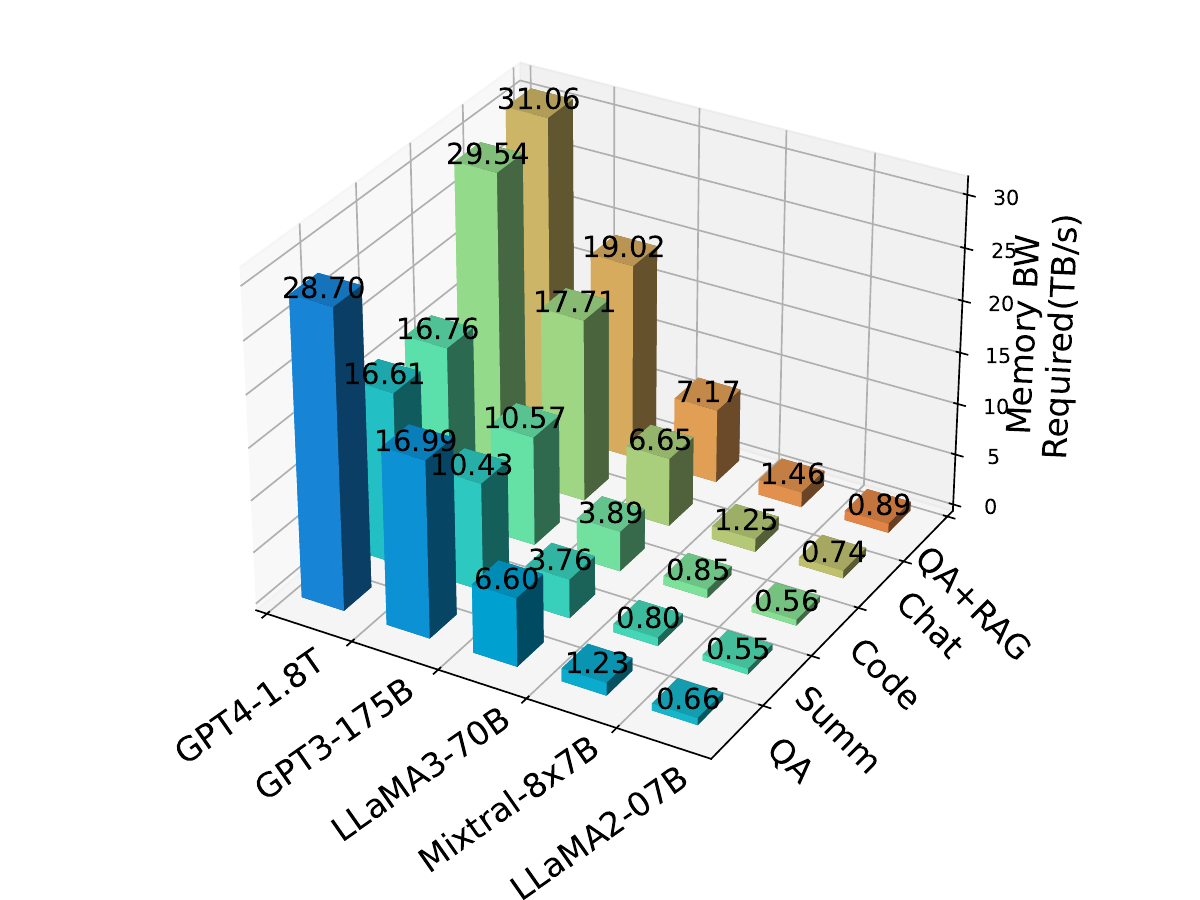}
        \text{(b) Memory Bandwidth(TB/s) requirement.}
        \label{fig:Memory_BW_requirement}
    \end{subfigure}
    \caption{Platform Scale requirements.}
\label{fig:platform_scale_requirement}
\end{figure}

\subsection{Platform Compute Requirement}
\label{subsec:platform_FLOPS_req}

%
%
Given that the prefill stage is predominantly compute-bound, the platform prerequisites for TFLOPs are determined by prefill stage across all scenarios. 
%
\autoref{fig:platform_scale_requirement}(a)
shows the platform-level compute TFLOPs required to meet the requirements in \autoref{tab:usecases} for various models.
\\

The compute TFLOPS is contingent upon both the model's dimensions and the quantity of input tokens. In cases where there are lenient expectations regarding Time to First Token (TTFT), the required TFLOPS also decreases. 
When using RAG for question answering, the increased prompt size causes the TFLOPS requirement to go up by 5.41x across all models. 
%

\noindent\fbox{%
  \parbox{23em}{
  \textit{Key Takeaways:} \\
    $\bullet$ Fixed-usecase, TFLOPS$_{Req} \propto O(\frac{Model Size + KV cache}{TTFT})$. \\
    $\bullet$ Fixed-model, TFLOPS$_{Req} \propto O(\frac{B * \tau_p}{TTFT})$.
  }%
}

\subsection{Platform Memory Bandwidth Requirement}
\label{subsec:platform_BW_req}
%
%
The memory bandwidth required to satisfy TPOT constraints dictates the platform requirements.
%
%
\autoref{fig:platform_scale_requirement}(b)
shows the platform level memory BW required to meet the requirements for various workloads.
We assume the LLM inference with model weights and KV cache in FP8 format. Changing the dataformat would proportionally scale up or down the bandwidth requirements.
For GPT4, the memory bandwidth going from QA to RAG based QA only increased the requirement by 8\% even with context length becoming 10$\times$. This is owing to the large size of the model itself.

\vspace{1em}
\noindent\fbox{%
  \parbox{23em}{
  \textit{Key Takeaways:} \\
    $\bullet$ Fixed-usecase, BW$_{Req}\propto O(\frac{Active Model + KV cache}{TPOT})$.\\
    $\bullet$ Fixed-model, BW$_{Req} \propto O(\frac{B * (\tau_p + S_b * \tau_d)}{TPOT})$.}%
}




\section{Studies on Future AI Inference Platform Design}
\label{sec:platform_impact}

In this section, we demonstrate the usefulness of \tool in generating 
insights for future AI inference platform design.


    
    
    

\begin{figure}[!t]
    \centering
    \begin{subfigure}
        \centering
        \includegraphics[width=25em]{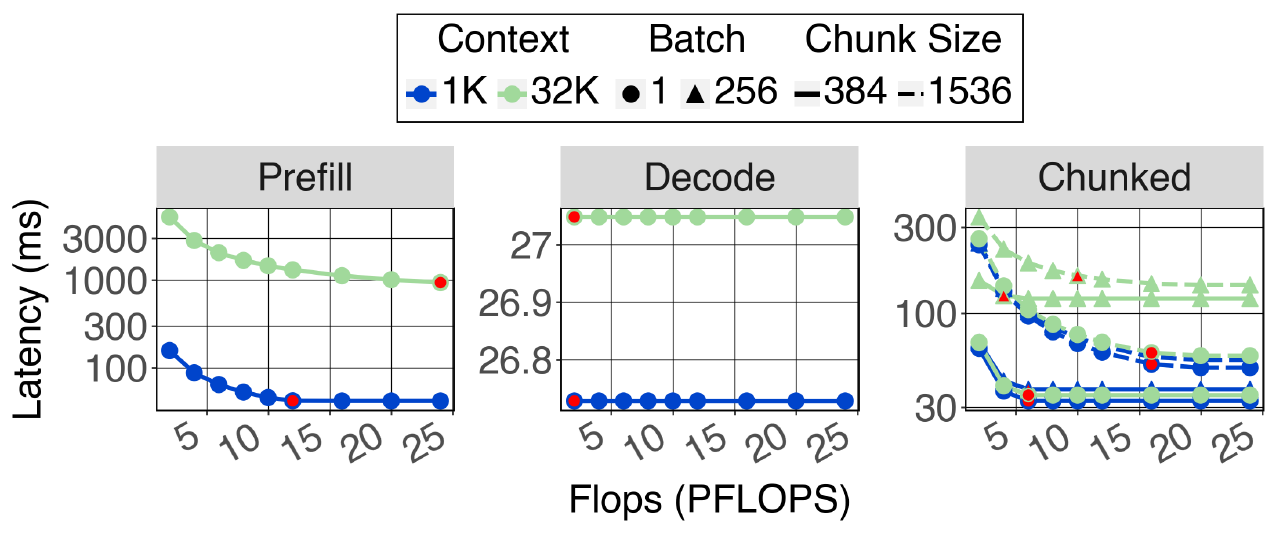}
        \text{(a) TFLOPS Scaling}
        \label{fig:TFLOPS_Scaling}
    \end{subfigure}
    \begin{subfigure}
        \centering
        \includegraphics[width=25em]{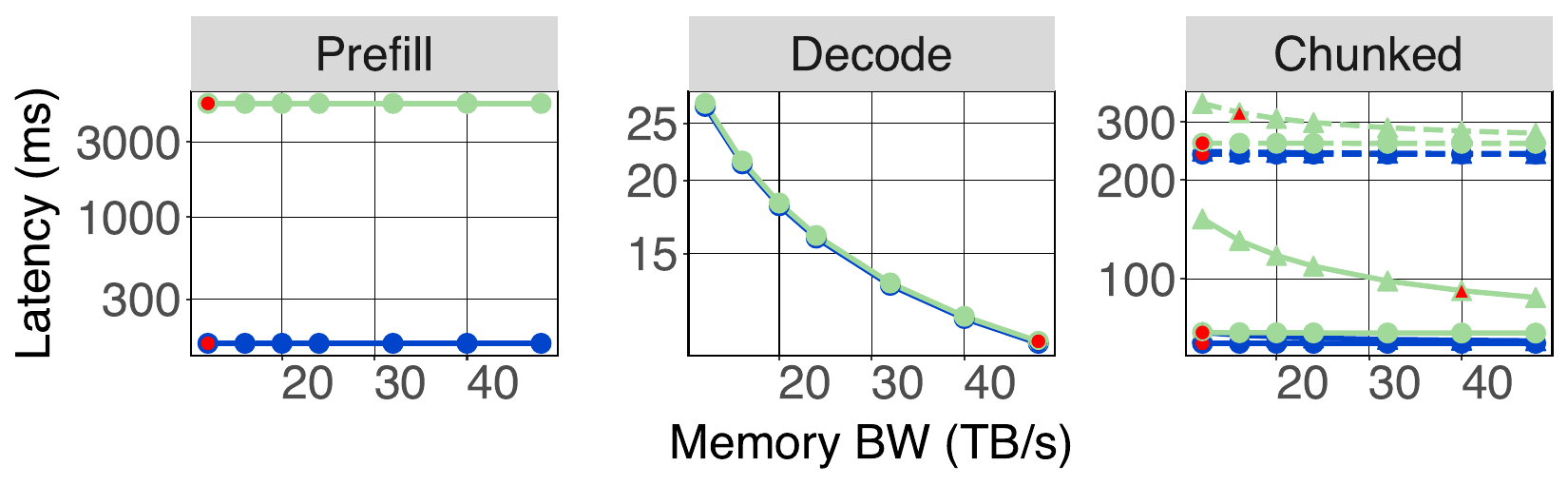}
        \text{(b) Memory Bandwidth Scaling}
        \label{fig:MemBW_Scaling}
    \end{subfigure}
    \begin{subfigure}
        \centering
        \includegraphics[width=25em]{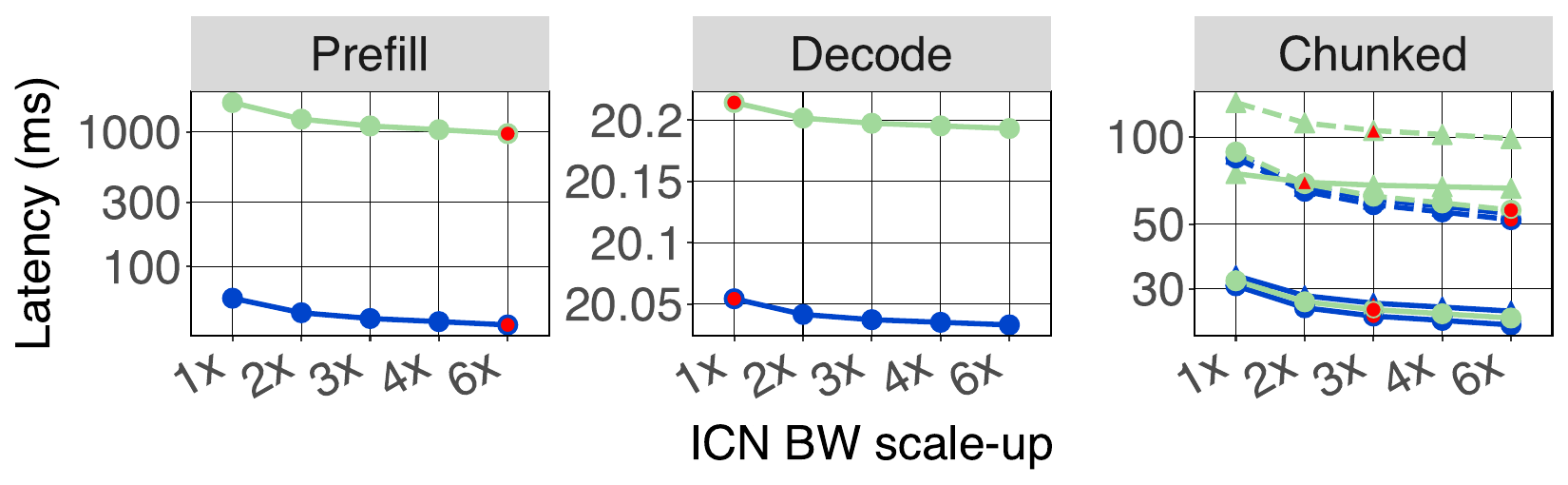}
        \text{(c) Interconnect Network Bandwidth Scaling}
        \label{fig:ICN_BW_Scaling}
    \end{subfigure}
    \begin{subfigure}
        \centering
        \includegraphics[width=25em]{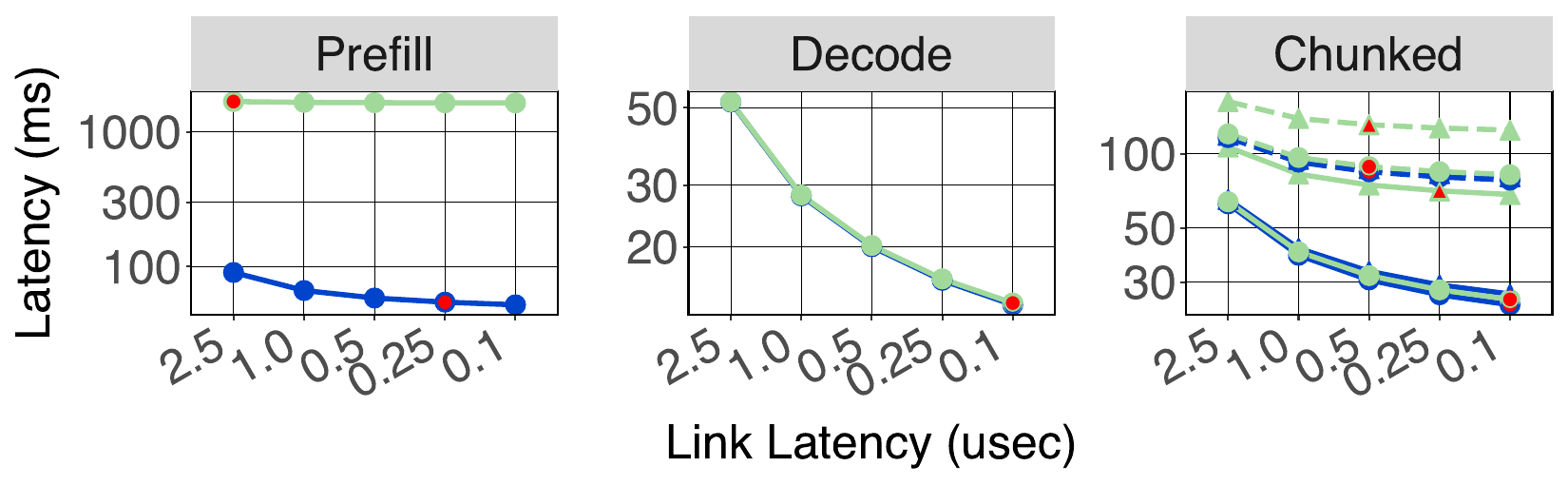}
        \text{(d) Interconnect Network Link Latency Scaling}
        \label{fig:ICN_LL_Scaling}
    \end{subfigure}
    \caption{Impact of scaling individual HW  characteristics on runtime of different stages. The \textcolor{red}{red} points in the plot represent the optimal points for the stage and workload. The fixed NPU parameters are FLOPs=2PFLOPs, Memory=360 GB @ 12 TB/s and ICN link with 500 us link latency and 1.8 TB/s bandwidth.
    }
    \label{fig:platform_scaling}
\end{figure}

\subsection{Case Study I: Scaling HW Characteristics}
\label{subsec:scaling_platform}
We study how improving different HW characteristics impacts the latency of different stages.
For all our experiments, we assume a hypothetical Dense-GQA-5T parameter model defined in \autoref{tab:model_parameters}.
%
We explore four key hardware characteristics: TFLOPS, Memory BW , ICN link bandwidth, ICN link latency.
To understand the effect of each characteristic on LLM inference performance, we vary each parameter in an isolated fashion. 
We vary the context length $\in \{1k, 32k\}$, and for running with chunking; we vary the chunk size $\in \{384, 1536\}$ and batch size $\in \{1, 256\}$. 
\autoref{tab:platform_scaling_summary} summarizes this case study's findings and \autoref{fig:platform_scaling} quantifies the observed trends.

\subsubsection{TFLOPS Scaling}

We vary the platform TFLOPS to increase the compute-to-memory BW ratio(C:M ratio). 
A higher C:M ratio benefits the operators with large arithmetic intensity due to additional compute units.
The prefill stage, especially with long context, gets a very good reduction in latency as C:M increases. For smaller context, the improvement stops after C:M reaches 2000. 
The decode doesn't improve as expected due to all layers being memory-bound.
For chunking, the larger chunk sizes with fewer decode batches has the highest improvement with an increase in the C:M ratio.
With more decode batches, or the smaller chunk size, there is only limited gain from the TFLOPs increase.

\subsubsection{Memory Bandwidth Scaling}
Next, we increase the memory bw of each NPU in the AI platform, thus reducing the C:M ratio.
The prefill stage has no benefit from memory BW boost since it is completely compute-bound.
The decode stage latency drops proportionally to memory BW increase since the decode stage is traditionally memory-bounded.
For the chunking, the primary benefit is for scenarios where decode batches have accumulated and may constitute a significant portion of the overall runtime.
For all other cases, there is no improvement in the chunk processing latency due to memory bandwidth improvement.

\subsubsection{ICN Bandwidth Scaling}
Our studies were done with a 32-NPU AI platform. Thus, communication latency composes a significant portion of the total latency. To further reduce the communication latency component, we scale the ICN BW.
We see that the prefill stage benefits tremendously from bandwidth scale-up since the large message sizes in the prefill are primarily ICN bandwidth bound.
Decode generally has much smaller message sizes, O($\approx$ 50-100 kBs); thus, bandwidth scale-up doesn't improve its performance.
With chunking, the message sizes are proportional to the chunk sizes. Thus, cases with large chunk-size benefit the most from the ICN BW increase. With chunk size, the benefit is limited since the collective might not be as significant a portion as other layers.

\subsubsection{ICN Link Latency Scaling}

Finally, we study the impact of platform interconnect network link latency; we reduce the $T_{link}$ remove 2.5 us to 0.1 us.
For prefill, the communication component in a large context is smaller, and thus, reducing link latency does not significantly impact the run-time but is helpful for smaller context workloads.
In the decoding process, link latency constitutes the majority of the total time involved. Therefore, a significant reduction in link latency leads to a notable decrease in decode latency. 
For chunking, reducing link latency has a more pronounced effect on smaller chunk sizes, while its impact on larger chunk sizes is minimal. 

\begin{table*}[!bhpt]
\centering
\begin{tabular}{|c|c|c|c|}
\hline
\rowcolor{gray!20} 
{\textbf{Characteristic}} & \textbf{Prefill Stage} & \textbf{Decode Stage} & \textbf{Chunked-Prefill} \\ \hline
\textbf{TFLOPS} & \begin{tabular}[c]{@{}c@{}}\textcolor{green!70!black}{$\uparrow$ Large} (long \\ context) \\ \textcolor{orange!80!black}{$\uparrow$ Small} (short \\ context)\end{tabular} & \makecell{\textcolor{red!70!black}{\ding{55}} (Memory-\\bound)} & \begin{tabular}[c]{@{}c@{}}\textcolor{green!70!black}{$\uparrow$ Large} (large chunks, \\ fewer batches) \\ \textcolor{orange!80!black}{$\uparrow$ Small} (small chunks, \\ more batches)\end{tabular} \\ \hline
\begin{tabular}[c]{@{}c@{}}\textbf{Memory} \\ \textbf{BW} \end{tabular} & \makecell{\textcolor{red!70!black}{\ding{55}} (Compute-\\bound)} & \textcolor{green!70!black}{$\uparrow$ Proportional} & \begin{tabular}[c]{@{}c@{}}\textcolor{green!70!black}{$\uparrow$ Only when decode} \\ \textcolor{green!70!black}{accumulates} \\ \textcolor{red!70!black}{\ding{55}} (otherwise)\end{tabular} \\ \hline
\textbf{ICN BW} & \makecell{\textcolor{green!70!black}{$\uparrow$ Large} (ICN- \\ bound large \\messages)} & \makecell{\textcolor{red!70!black}{\ding{55}} (Small  \\messages: \\50-100kBs)} & \begin{tabular}[c]{@{}c@{}}\textcolor{green!70!black}{$\uparrow$ Large chunks} \\ \textcolor{orange!80!black}{$\uparrow$ Small} (Small chunks)\end{tabular} \\ \hline
\textbf{\makecell{ICN Link \\Latency}} & \begin{tabular}[c]{@{}c@{}}\textcolor{orange!80!black}{$\uparrow$ Small} (short \\ context)\\  \textcolor{red!70!black}{\ding{55}} (large context)\end{tabular} & \textcolor{green!70!black}{$\uparrow$ Significant} & \begin{tabular}[c]{@{}c@{}}\textcolor{green!70!black}{$\uparrow$ Small chunks} \\ \textcolor{orange!80!black}{$\uparrow$ Small} (Large chunks)\end{tabular} \\ \hline
\end{tabular}
\caption{Improvement in LLM Inference at different stages as we scale different platform characteristics. 
}
\label{tab:platform_scaling_summary}
\end{table*}

\definecolor{scale_up_color}{RGB}{53, 49, 255} 
\definecolor{scale_out_color}{RGB}{0, 206, 250} 

\definecolor{SRAM_color}{RGB}{255, 69, 0}  
\definecolor{HBM_color}{RGB}{255, 182, 0} 

\begin{table*}[!thp]
\resizebox{\linewidth}{!}{%
\begin{tabular}{|c|c|c|c|c|c|}
\hline
\textbf{Configuration} & \begin{tabular}[c]{@{}c@{}}\textbf{Platform}\\ \textbf{Inspiration}\end{tabular} & \begin{tabular}[c]{@{}c@{}}\textbf{Compute/node}\\ \textbf{PFLOPS}\end{tabular} & \begin{tabular}[c]{@{}c@{}}\textbf{Memory per node}\\ (\textcolor{SRAM_color}{\textbf{SRAM}}/\textcolor{HBM_color}{\textbf{HBM}})\end{tabular} & \begin{tabular}[c]{@{}c@{}} \textbf{Network} \\ \textcolor{scale_up_color}{\textbf{Scale-up}} \textcolor{scale_out_color}{\textbf{Scale-out}}\end{tabular} & \begin{tabular}[c]{@{}c@{}}\textbf{Peak Platform}\\ \textbf{Power(kW)}\end{tabular} \\ \hline
\begin{tabular}[c]{@{}c@{}}Multiple\\ GPUs\end{tabular} & Nvidia-GB200~\cite{Comparin94:online} & 4.5 & \begin{tabular}[c]{@{}c@{}}\textcolor{SRAM_color}{128MB@40 TB/s}\\ \textcolor{HBM_color}{192GB@8 TB/s}\end{tabular} & \begin{tabular}[c]{@{}c@{}}\textcolor{scale_up_color}{Switch\{8\}@900GB/s}\\ \textcolor{scale_out_color}{Switch\{4\}@900GB/s}\end{tabular} & 57.2$^{\dagger}$ \\\hline
\begin{tabular}[c]{@{}c@{}}Single\\ SRAM wafer\end{tabular} & Cerebras-CS3~\cite{cerebras} & 125 & \begin{tabular}[c]{@{}c@{}}\textcolor{SRAM_color}{44GB@21 PB/s}\\ \textcolor{HBM_color}{12TB@14.6 TB/s}\end{tabular} & \begin{tabular}[c]{@{}c@{}}\textcolor{scale_up_color}{On-Wafer@214 PB/s}\\ \textcolor{scale_out_color}{-}\end{tabular} & 23 \\ \hline
\begin{tabular}[c]{@{}c@{}}Multiple\\ SRAM chips\end{tabular} & Groq-GroqChip~\cite{Groq}  & 0.75 & \begin{tabular}[c]{@{}c@{}}\textcolor{SRAM_color}{256MB@80 TB/s}\\ -\end{tabular} & \begin{tabular}[c]{@{}c@{}}\textcolor{scale_up_color}{FC\{64\}@3.2TB/s}\\ \textcolor{scale_out_color}{Ring\{16\}@256GB/s}\end{tabular} & 276.8$^{\star}$ \\ \hline
\begin{tabular}[c]{@{}c@{}}Multiple\\ ASICs\end{tabular} &Etched-Soho~\cite{Etched} & 45 & \begin{tabular}[c]{@{}c@{}}\textcolor{SRAM_color}{256MB@80 TB/s}\\ \textcolor{HBM_color}{192GB@8 TB/s}\end{tabular} & \begin{tabular}[c]{@{}c@{}}\textcolor{scale_up_color}{Switch\{8\}@900GB/s}\\ \textcolor{scale_out_color}{Switch\{4\}@900GB/s}\end{tabular} & 96$^{\ddagger}$ \\ \hline
\end{tabular}%
}
\caption{Platform architecture designs for comparison. Each of our platform's architectures is inspired by a current system architecture.
$^{\dagger}$ Power of 4 x DGX B200 platform(8xB200 GPU)\mbox{~\cite{Comparin94:online}}.
$^{\ddagger}$ Assumed power of a larger ASIC platform (NPUs have 10x flops of B200 GPU).
$^{\star}$Power of 16 GroqRack\cite{GroqRack}.
The architectures in the table are mapped to GenZ parameters defined in \autoref{sec:HW_acc_charac}, \autoref{sec:HW_platform_charac}. Column 3 is $FLOPS$, col 4  has BW$_{mem}$, Cap$_{mem}$, BW$_{sram}$, Cap$_{sram}$, col 5 has scaling dimension along with BW$_{link}$ for each network level.
}

\label{tab:platform_arch}
\end{table*}

\begin{figure*}[!bhtp]
    \centering
    \begin{subfigure}
    \centering
    \includegraphics[width=\linewidth]{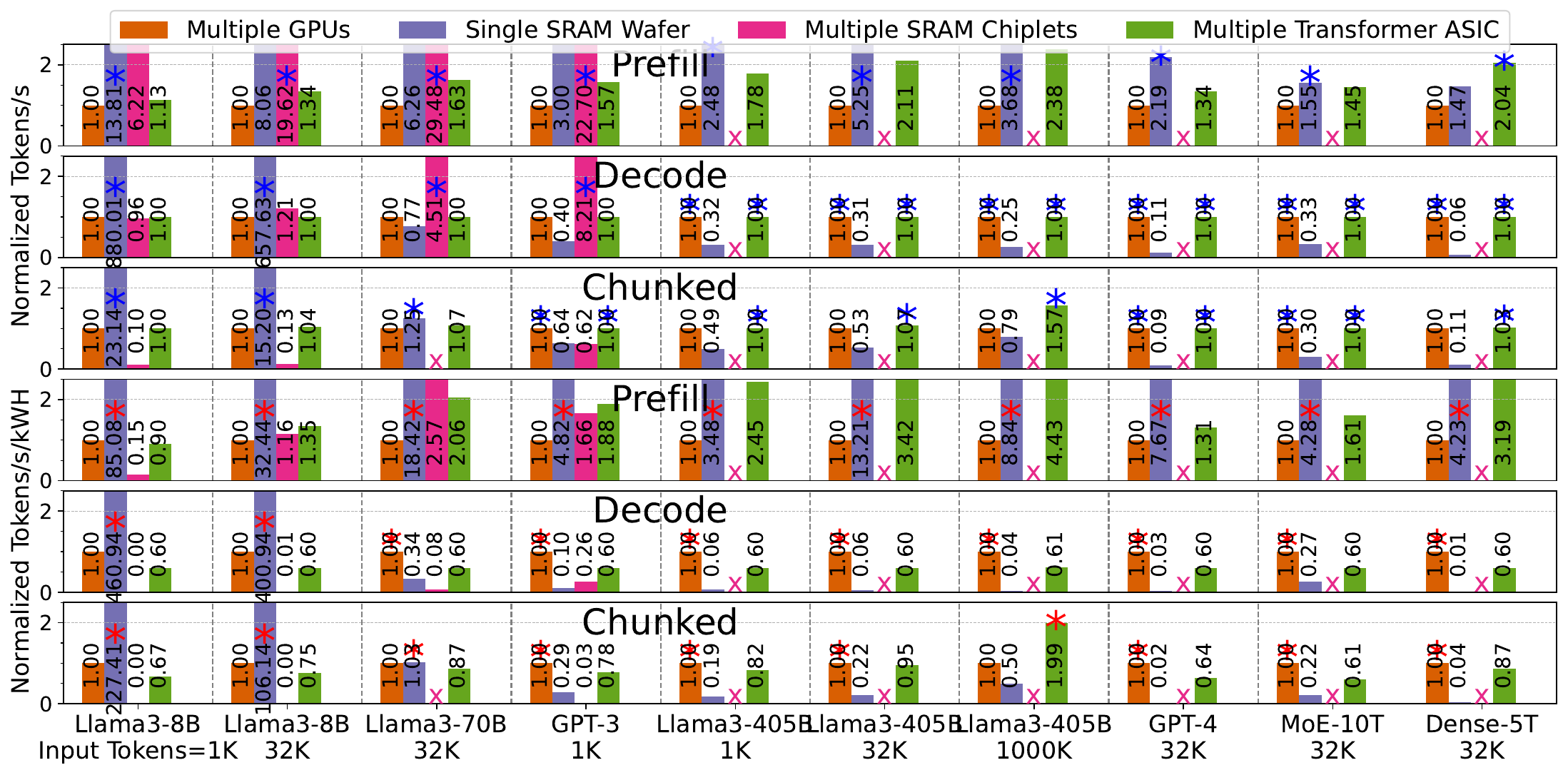}
    \label{fig:platform_comp_perf_kwh}
    \end{subfigure}
    \caption{Comparing normalized throughput and throughput/energy of different platform architectures across various workloads. `X' represents that the architecture is out of memory when running that workload. \textcolor{red}{\textbf{*}}/\textcolor{blue}{\textbf{*}} represents the platform architecture with the best performance. We run all workloads with batch size 4, $\tau_d = 1024$, and chunk size 512. }
    \label{fig:platform_comp}
\end{figure*}


\subsection{Case Study II: Comparing Diverse Platforms}
\label{subsec:c2_arch_comparision}

LLM inference throughput is heavily correlated to the characteristics of the AI platform used to serve LLMs. How to build next-generation LLMs is an open research question. Even commercially, industrial giants have hedged their bets on different platform architectures. We can categorize these architectural choices into four key buckets:

\noindent (1) \textbf{Multiple GPUs:} Traditional general-purpose SIMD or dataflow machines connected with a memory cache on-chip and connected to large memory banks.
Examples of these are GPUs~\cite{NVIDIABlackwellBox, NVIDIAH135:online, NvidiaH200,  amdMI300X,  intel_gaudi}, TPUs~\cite{tpu_pods, tpu-isca}, and other AI accelerators ~\cite{sambanova, AWSGravi55:online, ProdBrie73:online, HotChips65:online}.

\noindent (2) \textbf{Wafer-scale chips} with uber-fast on-wafer interconnect connecting cores and having very large on-chip SRAM, connected to large off-chip memory ~\cite{cerebras}.

\noindent (3) \textbf{Multiple SRAM chips:} Cluster of small chipset-based accelerators with large on-chip SRAM without any back-up memory.
~\cite{Groq,peng2023chiplet}.

\noindent (4) \textbf{Transformer-specific ASICs} with a very large number of compute cores with a small memory cache on-chip and connected to large memory banks~\cite{Etched}.

%
Using \tool, we compare four representative platform architectures defined in \autoref{tab:platform_arch} running various workloads for current models (8B, 70B, 405B) and future model architectures (5T, 10T).
Since the size and architecture of these platforms are completely different, there are different costs associated with each system. As we can't estimate the cost of all components, we use energy consumed as a proxy and Tokens/kWH as the comparison metric. We calculate the energy used in running workloads on different platforms. The energy consumption of each platform is modeled as a linear function of the utilization of its individual components ~\cite{AccelWattch}. We consider four main power components: Static/Idle, Compute, Memory, and Network. For each operator, the total energy is:

\begin{equation}
    E_{\text{op}} = T_{\text{op}} \times \left( P_{\text{static}} + P_{\text{C}} \cdot U_{\text{C}} + P_{\text{mem}} \cdot U_{\text{mem}} + P_{\text{icn}} \cdot U_{\text{icn}} \right)
\end{equation}

where $E_{\text{op}}$/$T_{\text{op}}$ are the operator energy/execution time, $P_{\text{static}}$  is the static power, and $P_{\text{C}}, P_{\text{mem}}, P_{\text{icn}}$ represent the peak power consumption of the compute, memory and network components, respectively. The terms $U_{\text{C}}, U_{\text{mem}}, U_{\text{icn}}$ denote their corresponding utilization factors for given operator.
We use $P_{\text{static}}:P_{\text{C}}: P_{\text{mem}}:P_{\text{icn}}::3:4:2:1$ \footnote{Future studies can use more fine-grained energy modeling simulators such as Accelergy~\cite{iccad_2019_accelergy}, AccelWatch~\cite{AccelWattch}.
}.

\autoref{fig:platform_comp} compares the normalized throughput and normalized throughput/energy of four platform architectures across various workloads and stages of LLM inference.
We run all workloads with batch size 4 and $\tau_d = 1024$. For the chunked stage, we use a chunk size of 512. For LLama3 \& GPT-3, we use TP=8 for GPUs and ASIC, TP=64, and PP=16 for SRAM chipsets. The use case indicates the model that was running and the input context size. 
For GPT-4, we use TP=32 for GPUs and ASIC.
A summary of their performance is as follows:
\circled{1} \textbf{GPUs:} Excel in most decode and chunked workloads when the model cannot fit into SRAM, benefiting from high aggregate memory bandwidth.
\circled{2} \textbf{Single SRAM Wafer:} Leads in prefill workloads across most use cases due to superior energy efficiency. Performs best in decode and chunked stages when the entire model and weights fit within SRAM.
\circled{3} \textbf{SRAM Chiplets:} Optimal when model fits on chips but are consistently outperformed by other architectures in perf/energy, primarily due to their high power consumption owing to their very large platform size O(100s chips).
\circled{4} \textbf{Transformer ASICs:} Thrive in high compute demands, especially for future larger models/large context lengths. 
%
\tcolorbox[colback=lightgray,colframe=black]

  $\bullet$ \textbf{Choosing platform architecture for Performance}: SRAM wafer/SRAM chips provide the best performance \textit{when the model fits on SRAM} due to superior on-chip memory BW. ASIC is best in the prefill stage for very large models. In decode and most chunked workloads, ASIC and GPU give similar performance due to similar memory bandwidth.\\
  \endtcolorbox
\tcolorbox[colback=lightgray,colframe=black]

  $\bullet$ \textbf{Choosing platform architecture for Perf/energy}: SRAM Wafer is best for prefill stage and decode/chunked when the entire model fits on SRAM as lower energy used for running a single chip compared to multiple racks of NPUs, GPUs are best for running the rest of the decode and chunked workloads due to lower energy consumption accored to denser ASIC chip. If the ASIC chip can run with similar or lower power than GPU, it would be the best choice for achieveing highest performance/kWH in future larger models.
  \endtcolorbox

  

\subsection{Case Study III: Exploring HBD Design Choices}
\label{subsec:c3_network_comparision}

As model sizes continue to scale, the minimum number of NPUs required to meet stringent Service Level Objectives (SLOs) has grown steadily. For instance, running GPT4, demands 64 H100 GPUs. Looking ahead, this baseline is poised to rise further, necessitating careful planning not just for NPU hardware but also for the accompanying network architecture—a challenge in its own right.

The concept of a high-bandwidth domain (HBD), defined as a group of NPUs interconnected via high-bandwidth scale-up links, plays a pivotal role here. Nvidia has progressively expanded the size of HBDs, scaling from 8 NPUs in the DGX H100 system to an impressive 72 NPUs in the GB200 NVL72. This case study delves into determining the optimal HBD size and designing effective interconnections between HBDs. To this end, we explore and compare various network architectures across a 256-NPU setup.
We use the same individual NPU with 9 PFlops compute and 256 GB HBM, providing 13.5 TB/s for all configurations.
\autoref{tab:network_specs} shows the bandwidth and link latency of different interconnect types that we consider for building the platforms. We keep the network topology fixed as 2 levels of switch followed by a ring for the third dimension.
Using these, we build five different network architectures shown in \autoref{tab:network_configs}. We ran models with TP=64 and PP=4.

\begin{figure}[t]
    \centering
    \includegraphics[width=1\linewidth]{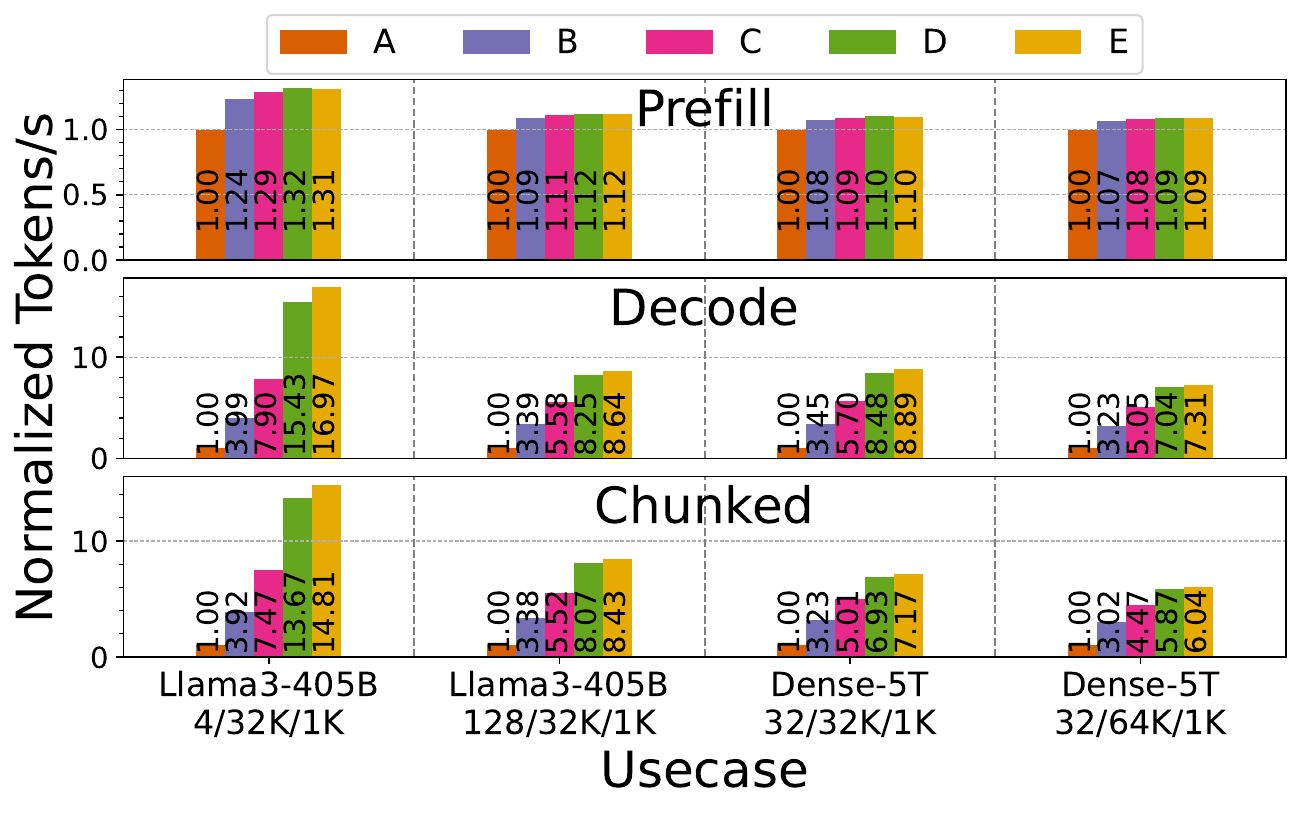}

    \caption{Comparing throughput for different network config.}
    \label{fig:two_dim_platform}

\end{figure}

\autoref{fig:two_dim_platform} shows the throughput of different network configurations across running different workloads on 256 NPUs.
Config D with all NPUs connected to a single HBD has the highest throughput but would also be the most costly to build.
In contrast, Config B, with 64 NPUs per HBD, gives a similar throughput for the prefill stage at a much lower cost.
Config E, with \textbf{64 NPUs per HBD connected via optical links as scale-out interconnect}, achieves throughput that is comparable to config D at a lower cost for all stages.

\definecolor{SL}{RGB}{128, 0, 128} 

\definecolor{IB}{RGB}{255, 140, 0}  
\definecolor{Optical}{RGB}{0, 100, 0} 

\begin{table}[ht]
    \centering
    \begin{tabular}{|c|c|c|c|c|}
        \hline
        & \textcolor{SL}{\textbf{High BW (SL)}} & \textbf{\textcolor{IB}{InfiniBand (IB)}} & \textcolor{Optical}{\textbf{Optical}} \\
        \hline
        \textbf{Latency (ns)} & \textcolor{SL}{500} & \textcolor{IB}{10000}  & \textcolor{Optical}{200} \\
        \hline
        \textbf{BW (GB/s)} & \textcolor{SL}{1800} & \textcolor{IB}{256} & \textcolor{Optical}{900} \\
        \hline
    \end{tabular}
    \caption{Link latency and bandwidth for different interconnects. High bandwidth(SL) is a scale-up link similar to NVlink, UAlink, etc. ~\cite{optical_latency, optim_collective, Optical_icn_BW, intel_gaudi, nvidia_sharp, infiniband:online, ualink:online, sip-ml, Li_2020}.}
    \label{tab:network_specs}
\end{table}

\begin{table}[ht]
\centering

\begin{tabular}{|c|c|c|c|} 
\hline
\textbf{Config} & \textbf{NPU Counts} & \textbf{ICN Type} & \textbf{HBD Size}  \\ 
\hline
\textbf{A}      & 8, 8, 4             & \textcolor{SL}{SL}, \textcolor{IB}{IB}, \textcolor{IB}{IB}        & 8                  \\ 
\hline
\textbf{B}      & 8, 8, 4           & \textcolor{SL}{SL}, \textcolor{SL}{SL}, \textcolor{IB}{IB}        & 64                 \\ 
\hline
\textbf{C}      & 8, 16, 2            & \textcolor{SL}{SL}, \textcolor{SL}{SL}, \textcolor{IB}{IB}        & 128                \\ 
\hline
\textbf{D}      & 8, 8, 4             & \textcolor{SL}{SL}, \textcolor{SL}{SL}, \textcolor{SL}{SL}        & 256                \\ 
\hline
\textbf{E}      & 8, 8, 4             & \textcolor{SL}{SL}, \textcolor{SL}{SL}, \textcolor{Optical}{Optical}   & 64                 \\
\hline
\end{tabular}
\caption{Comparison of different network configurations}
\label{tab:network_configs}

\end{table}

\subsection{Case Study IV: Exploring Different NPU Microarchitectures and Offloading Choices}
\label{subsec:c4_uarch_casestudy}

\begin{figure}
    \centering
    \includegraphics[width=1\linewidth]{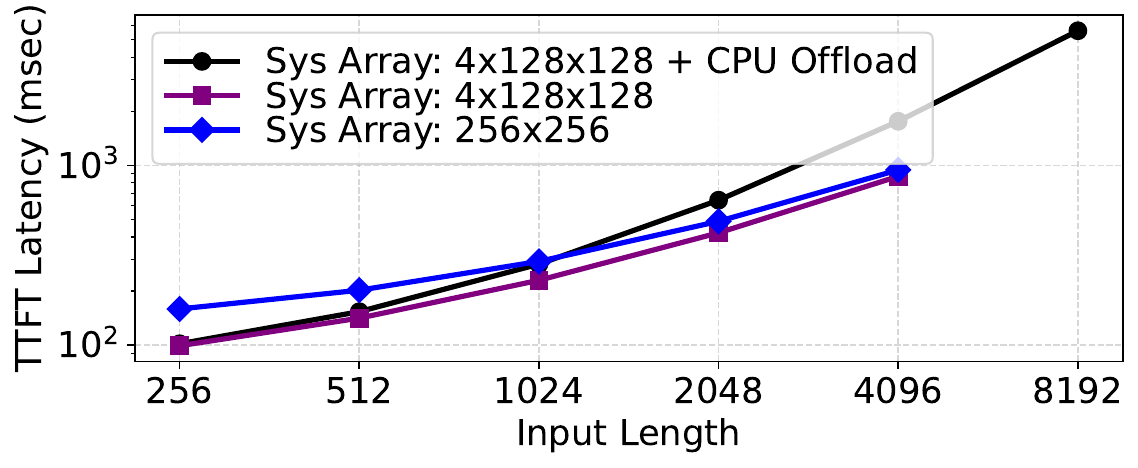}
    \caption{Comparing different micro-architectures choices and compute offloading running LLama-3-8B prefill stage.}
    \label{fig:uarch_modelling}
\end{figure}

We compare systems with identical platform architectures but differing NPU microarchitectures, using SCALE-sim. The NPUs employ a systolic array with weight-stationary dataflow and spatial mapping. Keeping the total MAC units constant, we evaluate three configurations:
(A) A single 256×256 systolic core.
(B) Four 128×128 cores.
(C) Four 128×128 cores with CPU offloading for MHA (Logit + Softmax + Attend) and KV cache storage. CPU has 8 TOPS and GPU to CPU link of 128 GB/s using PCIe.
All NPUs are connected to a single 16GB HBM3e stack operating at 1.2 TB/s.
\autoref{fig:uarch_modelling} compares the prefill latency of these systems running LLaMA 3-8B (BF16) across varying input context sizes. System B achieves the lowest latency due to finer-grained kernel scheduling. System C, despite performance degradation from CPU offloading as the KV cache grows, can handle longer sequences, unlike Systems A and B, which are limited by fixed memory size.
This analysis highlights \toolns's capability to model diverse microarchitectures and offload compute. 

\subsection{Extreme Scale Requirements: AI assistant}
\label{sec:Extreme_scale}

Examining the trajectory of model scaling over the past couple of years suggests an ongoing trend of increasing model size\cite{informationisbeautiful}. Moreover, there has been a steady rise in the maximum context lengths of large language models (LLMs).\cite{humanloop}.  
For instance, the GPT-4-Prompts dataset, comprising multi-turn conversational prompts, already features prompts of up to 900k tokens.\cite{gpt4Prompts}. 
Recently introduced, Google Gemini can accommodate prompts of up to 1M tokens in production\cite{Gemini}.
In this section, we delineate the platform requirements for serving these forthcoming LLMs with extensive context lengths and juxtapose them with the current state-of-the-art standards.

We envision an enormous 10T parameter Super-LLM model as the futuristic model from \autoref{tab:model_parameters}. 
Our aim is to employ the Super-LLM as a standard AI assistant capable of engaging in \textbf{real-time conversations with humans}, furnishing answers on diverse topics, and retaining recollections of past interactions.
Hence, supporting large context windows is imperative for AI assistants. 
This workload can be characterized as $S_b = 4, \tau_p = \text{Variable}, \tau_d=2000$.
Given its role as a personal assistant, we would operate it with a batch size of 1. For seamless conversation, we assume it can generate output at the rate at which humans read or listen, which amounts to 300 words per minute\cite{BRYSBAERT2019104047}.

\begin{figure}[t!]
    \centering
    \begin{subfigure}
        \centering
        \includegraphics[width=11.5em]{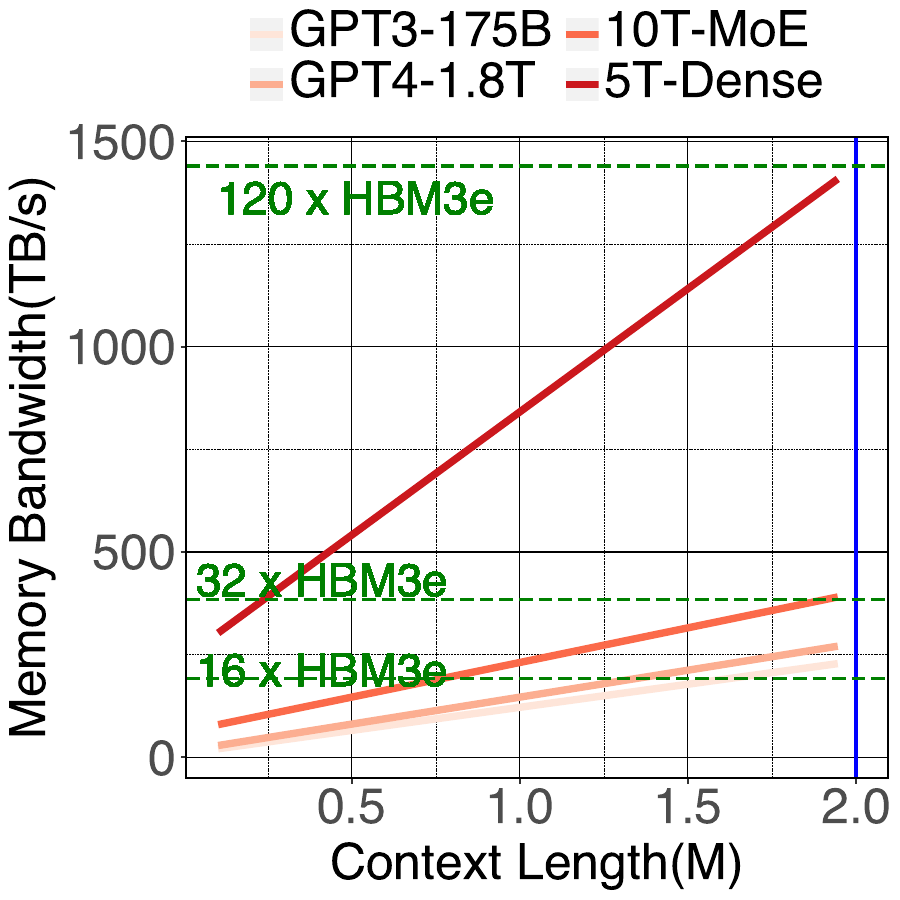}
        \label{fig:extreme_memory_bw_req}
    \end{subfigure}
    \begin{subfigure}
        \centering
        \includegraphics[width=11.5em]{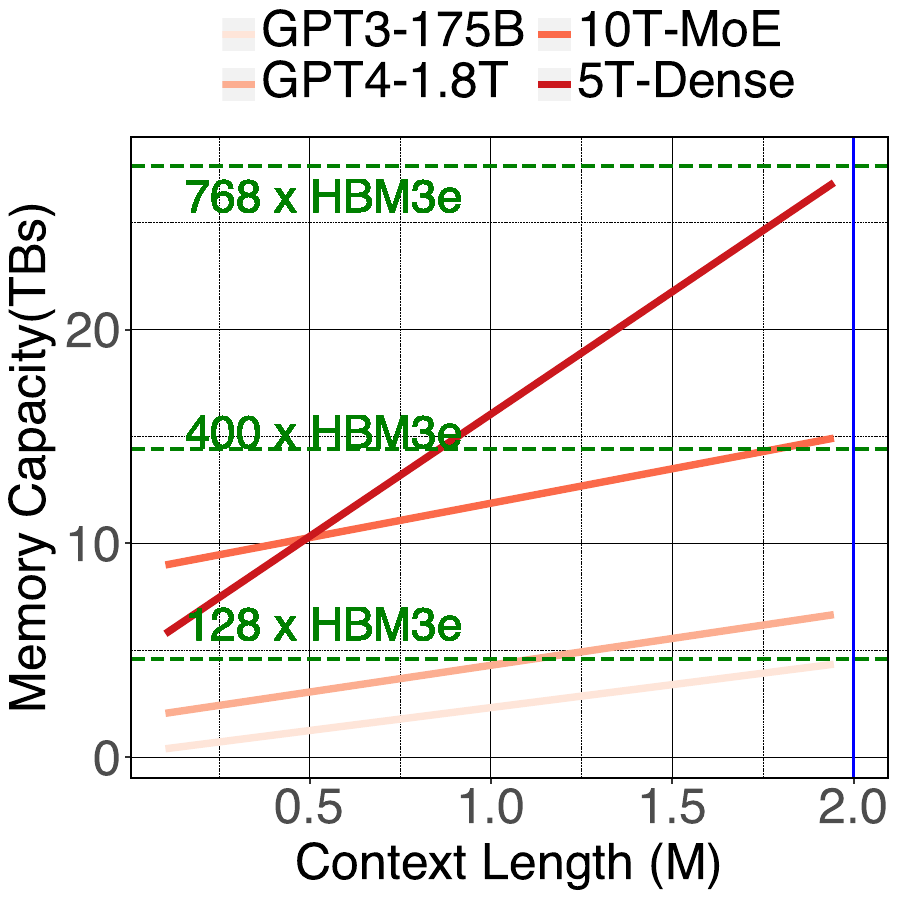}
        \label{fig:extreme_memory_cap_req}
    \end{subfigure}
    \caption{Platform Scale requirements supporting a real-time conversational AI personal assistant.}
    \label{fig:ai_assistant_req}
\end{figure}

For real-time conversations, the KV cache would grow linearly with context, leading to higher memory capacity and bandwidth requirements.
\autoref{fig:ai_assistant_req} shows the memory bandwidth and capacity required for using the super LLM as context lengths scale to 2M tokens. 
To gauge its feasibility, we compare it against the current SOTA memory, HBM3e, with 1.2TB/s BW and 36 GB memory capacity per stack.
We require 40 TB/s memory bandwidth to make a super-LLM AI assistant, which equates to roughly 32 HBM3e stacks.
On the other hand, it would require around 15 TB of memory capacity, equating to 400 HBM3e stacks.

\vspace{1em}
\fbox{%
  \parbox{0.9\columnwidth}{
\textit{  Key Takeaway: }\\
  $\bullet$ The platform memory capacity seems to be a more intensive bottleneck than the memory bandwidth.\\
  $\bullet$ Memory BW is within a reasonable range due to the sparse activation of the model. The growth of memory capacity requirement seems unsustainable.
  }%
}





\section{Related Works}
\label{sec:related_works}

\textbf{LLM inference serving and analysis:} We are seeing a massive uprising of works related to LLM inference serving. There are works on optimizing the batching~\cite{yu2022orca, holmes2024deepspeed, agrawal2023sarathi}, memory optimization~\cite{dao2022flashattention, pagedAttention} and scheduling~\cite{sheng2023flexgen, fu2024break}. 
Various articles~\cite{databricks,rel_art1,rel_art2,rel_art3,rel_art4,rel_art5,rel_art6} 
provides metrics, analysis, insights, and best practices for LLM inference performance on current hardware systems. 
Various works provided a survey ~\cite{yuan2024llm, kim2023full, chittyvenkata2023survey, minaee2024largelanguagemodelssurvey, fan2023largelanguagemodelssoftware} of transformer inference and various optimizations in existing transformer architecture on current hardware systems. 
%

\textbf{Tools for AI Platform Modeling:}  Table \ref{tab:related_works} highlights key
differences between \tool and other performance models for
LLMs. 
%
%
%
 Simulators like ASTRA-sim~\cite{astrasim}, MadMax ~\cite{hsia2024mad} and vTrain ~\cite{bang2023vtrain} focus on communication optimizations for distributed \textit{training} and do not model different LLM architectures or LLM model optimizations on a distributed inference.
Tools like LLM-viewer~\cite{yuan2024llm} provide ideal roofline estimation.
Vidur~\cite{agrawal2024vidurlargescalesimulationframework} is an inference system simulator that focuses on studying the impact of different scheduling techniques on current hardware.
LLMServingSim~\cite{cho2024llmservingsimhwswcosimulationinfrastructure} provides a framework to compare scheduling algorithms.
LLM-Compass~\cite{zhang2023hardware} does DSE to optimize and generate ASIC configuration(Systolic arrays, cache sizes) for running dense models.
There is a lack of a single tool/framework in the community that can help us study different LLM model architectures combined with the latest optimizations on distributed NPU platforms. 




%
%

\section{Extensions and Future Directions}
\label{sec:future_direction}
Our work intentionally abstracts low-level microarchitectural details by encapsulating them as efficiency factors and using external tools for refined performance estimates.
%
We also provide linear energy estimation and aggregated batching support, future works can extend these capabilities with advanced modelling (e.g. additional scheduling modules to simulate disaggregated /heterogeneous serving).

\section{Conclusions}
\label{sec:conclusion}


We introduce GenZ, a framework with an indispensable capability
of navigating the intricate design space of LLM inference, quantifying the interplay between various model and system-level optimizations, and steering the development of future AI platforms.
We also show four key studies to present a roadmap for improving next-generation AI platform characteristics and selecting the appropriate architectural paradigms for running LLM inference. The source code is available at \href{https://github.com/abhibambhaniya/GenZ-LLM-Analyzer}{GitHub}. \tool can also be tried out on its \href{https://genz-llm-analyzer.streamlit.app/}{website} without any setup on your web browser.

\section{Acknowledgment}
This work was supported in part by CoCoSys, one of seven centers in JUMP 2.0, a Semiconductor Research Corporation (SRC) program sponsored by DARPA. We thank the anonymous reviewers for their insightful feedback, which significantly improved the quality of this work.
We are especially grateful to Amir Yazdanbakhsh for his valuable discussions and feedback, which helped shape the foundation of this research.
We also thank Arijit Raychoudhury for his thoughtful discussions that were instrumental in realizing the full potential of this work.

\bibliographystyle{IEEEtran}
\bibliography{refs}

\end{document}